\documentclass{aa}  

\usepackage{graphicx}
\usepackage[varg]{txfonts}
\usepackage{amsmath}
\usepackage{color}
\usepackage{xcolor}
\usepackage{hyperref}
\usepackage{siunitx}
\hypersetup{
    unicode=true,           
    pdftoolbar=true,        
    pdfmenubar=true,        
    pdffitwindow=true,      
    pdftitle={The complex dynamical past and future of CzeV343},
    pdfauthor={Pejcha et al.},
    pdfkeywords={},         
    pdfnewwindow=true,      
    colorlinks=true,        
    linkcolor=blue,         
    citecolor=blue,         
    filecolor=gray,         
    urlcolor=blue           
}

\makeatletter
\renewcommand*\aa@pageof{, page \thepage{} of \pageref*{LastPage}}
\makeatother

\newcommand{\msun}{\ensuremath{M_\odot}}
\newcommand{\rsun}{\ensuremath{R_\odot}}
\newcommand{\feh}{\ensuremath{\text{[Fe/H]}}}

\begin{document}

   \title{The complex dynamical past and future of double eclipsing binary CzeV343: misaligned orbits and period resonance}
    \titlerunning{The complex past and future of CzeV343}
    \authorrunning{Pejcha et al.}
   \author{Ond\v{r}ej Pejcha\inst{\ref{utf}}
        \and
       Pavel Caga\v{s}\inst{\ref{utf},\ref{bso},\ref{sphe}}
          \and
          Camille Landri\inst{\ref{utf}}
          \and
          Michael M. Fausnaugh\inst{\ref{mit}}
          \and
          Gisella De Rosa\inst{\ref{stsci}}
          \and
          Jose L. Prieto\inst{\ref{udp},\ref{mil}}
          \and
          Zbyn\v{e}k Henzl\inst{\ref{henzl},\ref{sphe}}
          \and
          Milan Pe\v{s}ta\inst{\ref{utf}}
          }

   \institute{Institute of Theoretical Physics, Faculty of Mathematics and Physics, Charles University, V Hole\v{s}ovi\v{c}k\'{a}ch 2, Praha 8, 180 00, Czech Republic, \email{pejcha@utf.mff.cuni.cz} \label{utf}
   \and
   BSObservatory, Modrá 587, 760 01 Zlín, Czech Republic, \email{pc@bsobservatory.org} \label{bso}
          \and
          Department of Physics and Kavli Institute for Astrophysics and Space Research, Massachusetts Institute of Technology, Cambridge, MA 02139, USA \label{mit}
          \and
          Space Telescope Science Institute, 3700 San Martin Drive, Baltimore, MD 21218, USA \label{stsci}
          \and
          Núcleo de Astronomía, Universidad Diego Portales, Av. Ejército 441, Santiago, Chile \label{udp}
          \and
          Millennium Institute of Astrophysics, Santiago, Chile \label{mil}
          \and
          Hv\v{e}zdárna Jaroslava Trnky ve Slaném, Nosa\v{c}ická 1713, Slaný 1, 274 01, Czech Republic \label{henzl}
          \and 
        Variable Star and Exoplanet Section, Czech Astronomical Society, Fričova 298, 251 65 Ondřejov \label{sphe}
             }

   \date{Received 2022}

 
  \abstract{CzeV343 (=V849 Aur) was previously identified as a candidate double eclipsing binary (2+2 quadruple), where the orbital periods of the two eclipsing binaries ($P_A \approx 1.2$\,days and $P_B \approx 0.8$\,days) lie very close to 3:2 resonance. Here, we analyze $11$ years of ground-based photometry, 4 sectors of TESS 2-minute and full-frame photometry, and two optical spectra. We construct a global model of our photometry, including apsidal motion of binary $A$ and light-travel time effect (LTTE) of the mutual outer orbit, and explore the parameter space with Markov Chain Monte Carlo. We estimate component masses for binary $A$ ($1.8+1.3\,\msun$) and binary $B$ ($1.4+1.2\,\msun$). We identify pseudo-synchronous rotation signal of binary $A$ in TESS photometry. We detect apsidal motion in binary $A$ with a period of about $33$ years, which is fully explained by tidal and rotational contributions of stars aligned with the orbit. The mutual orbit has a period of about $1450$\,days and eccentricity of about $0.7$. The LTTE amplitude is small, which points to low inclination of the outer orbit and a high degree of misalignment with the inner orbits. We find that when apsidal motion and mutual orbit are taken into account the orbital period resonance is exact to within $10^{-5}$ cycles/day. Many properties of CzeV343 are not compatible with requirements of the 3:2 resonance capture theory for coplanar orbits. Future evolution of CzeV343 can lead to mergers, triple common envelope, double white dwarf binaries, or a Type Ia supernova. More complex evolutionary pathways will likely arise from dynamical instability caused by orbital expansion when either of the binaries undergoes mass transfer. This instability has not been so far explored in 2+2 quadruples.}


   \maketitle
%

\section{Introduction}

Double eclipsing binaries (DEBs) are quadruple (2+2) stellar systems composed of two eclipsing binaries on a mutual orbit. The number of known DEBs has recently tremendously increased and the total sample of DEB candidates currently includes approximately 300 objects \citep{pawlak13,zasche19,zasche22,kostov22}. Relative frequency of DEBs on narrow orbits remains unknown, but DEBs on wide orbits are approximately 7 times more frequent than what would be expected from random pairings \citep{fezenko22} and in general quadruples of 2+2 hierarchy are over-represented among multiple stellar systems \citep{tokovinin14}.

In most cases, the mutual orbit is unresolved in optical ground-based imaging and the orbital parameters are unconstrained. For some objects, it was possible to constrain the mutual orbit with light travel time effect (LTTE) or with dynamical perturbations to the inner orbits. For example, \citet{zasche20} estimated outer orbital period in CzeV1731 to about 34 years. Most of DEBs characterized so far are compatible with aligned inner and outer orbits. \citet{zasche13} claimed that V994~Her has high inclination between the inner binaries and the mutual orbit, but later update by \citet{zasche16} showed that angular momenta of all three orbits lie in the plane of the sky. \citet{kostov21} found that all three orbits of TIC~454140642 are very well aligned and \citet{borkovits21} estimated that orbits in BG~Ind are closely aligned. Although the DEB statistics is small, we would expect at least some objects with misaligned orbits similarly to what was found by \citet{borkovits15} for stellar triples.

Relative frequency of different evolutionary pathways possible in 2+2 quadruples can differ from analogous triples, where one of the binaries is replaced by a single star. Completely new evolutionary pathways are also possible. \citet{pejcha13} performed direct integration of 2+2 quadruple orbits in the context of von Zeipel--Lidov--Kozai cycles acting on both inner binaries and found increased frequency of chaotic interactions, high-eccentricity encounters, collisions, and other interesting phenomena. This finding was confirmed with a secular model by \citet{vokrouhlicky16}. Later, \citet{hamers17} applied the secular model to hot Jupiters and \citet{fang18} to white dwarf mergers and Type Ia supernovae. Dynamics of 2+2 quadruples was further developed by \citet{hamers19}, \citet{hamers21}, \citet{fragione19}, \citet{liu19}, and \citet{vynatheya22} in the context of Type Ia supernovae, and neutron star and black hole mergers.

There are a number of open questions connected with the formation and evolution of 2+2 quadruples. One particularly striking feature is that in many DEBs the orbital periods of the two inner binaries, $P_A$ and $P_B$, appear in or near integer resonances. Unlike various types of dynamical perturbations between the orbits that are commonly studied in multiple stellar systems, period resonances are by principle absent in the dynamics of triples and are unique to 2+2 quadruples. To our knowledge, the first suggestion of a pair of interacting binaries with period ratio 3:2 was made by \citet{ofir08} for BI 108, a massive stellar system in the Large Magellanic Cloud \citep{kolaczkowski13}. \citet{cagas12} presented the discovery of CzeV343 (=V849 Aur), which was classified as a DEB composed of two detached binaries with periods of $P_A \approx 1.2$ and $P_B \approx 0.8$ days and period ratio\footnote{Throughout this paper we label with $A$ the binary with longer period and binary $B$ the binary with shorter period. This convention is opposite to the one used by \citet{tremaine20}.} very close to 3:2 resonance. V994~Her is somewhat farther from the 3:2 resonance, but its physical properties and parameters of the mutual orbit are relatively well known \citep{lee08,zasche13,zasche16}.  

\citet{zasche19} collected DEBs known at that time and claimed an excess of DEBs with period ratios of 3:2 and 1:1. Other DEBs were identified very close to other resonances of small integers such as 5:3 and 4:1 \citep{zasche22_4,zasche22}. More recently, \citet{kostov22} did not find any excess of period resonances in a sample of 97 DEBs identified in the data from the TESS satellite. The discrepancies in the published resonance statistics could arise due to different detection methods. So far, most of the DEB discoveries significantly rely to some extent on human factor, which might bias the detection efficiency. For example, near resonance of the two orbital periods can create light curve patterns that are more easily spotted by the human eye, which might lead to higher detection fraction near such period ratios. Addressing the issue of resonance statistics is beyond the scope of this work, but we note that in many cases it is not even clear how to quantify the distance of the period ratio from a resonance and how to reliably identify resonant DEBs. 

There are only two works addressing the theory of resonant orbits in quadruple stars.  \citet{breiter18} used canonical perturbation methods to construct a secular resonant model of coplanar binaries in 1:1 resonance. They concluded that the capture to 1:1 resonance is unlikely. \citet{tremaine20} extended the analysis to 3:2 and 2:1 resonances of coplanar orbits, where the resonant conditions include frequency of the mutual orbit and the rate of apsidal motion of one of the binaries. \citet{tremaine20} also derived several conditions on successful capture that include relative rate of tidal and orbital evolution of the individual binaries, and the eccentricities of the inner orbits and their evolution. \citet{tremaine20} concluded that CzeV343 and BI 108 are likely captured in the 3:2 resonance, while V994~Her is not, however, the observed eccentricities in some of these systems remain too high to be compatible with resonant capture.

In this paper, we revisit CzeV343, which was one of the first objects suggested to populate the 3:2 period resonance by \citet{cagas12}. Our goal is to characterize the properties of the system in light of new data and to evaluate how closely CzeV343 satisfies theoretical resonant conditions of \citet{tremaine20}. In Section~\ref{sec:observations}, we present new ground-based and TESS photometry as well as two spectra. In Section~\ref{sec:analysis}, we describe the modeling procedure for photometric and spectroscopic data. In Section~\ref{sec:results}, we present our results on apsidal motion, mutual orbit, rotation periods, and period resonance. In Section~\ref{sec:disc}, we discuss implications of our constraints on resonant capture and future evolution of the system. We summarize our findings in Section~\ref{sec:conc}.

\section{Observations}
\label{sec:observations}

\begin{figure*}
    \centering
    \includegraphics[width=\textwidth]{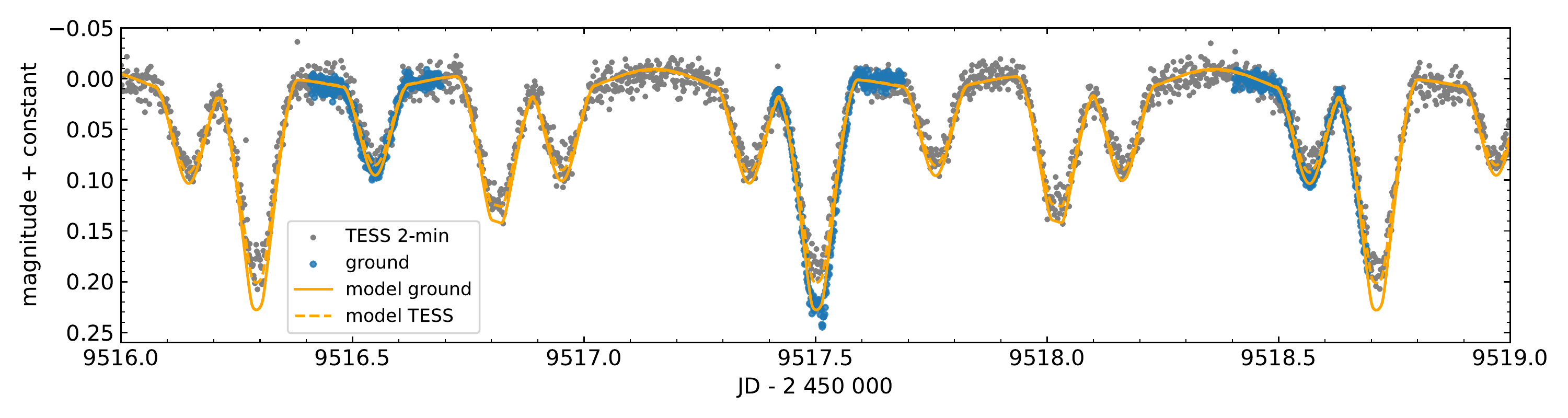}
    \caption{Section of light curve of CzeV 343 using our ground-based  (blue points) and TESS 2-min (gray points) measurements. As part of our model, both datasets were aligned vertically to match the double eclipsing binary light curve models, which are shown separately for the two datasets (solid and dashed orange lines).}
    \label{fig:lc_comp}
\end{figure*}

In this section, we discuss the properties and reduction process of our observations of CzeV343. We present ground-based photometry (Sect.~\ref{obs:ground}), TESS 2-min photometry (Sect.~\ref{obs:tess}), TESS full-frame images (Sect.~\ref{obs:ffi}), and two optical spectra (Sect.~\ref{obs:spectra}).

\subsection{Ground-based data}
\label{obs:ground}

We have been monitoring CzeV343 continuously since its discovery. Our instrumentation is similar to what was used by \citet{cagas12}, but virtually all components of the observing setup have been upgraded over time. The \SI{0.25}{m} telescope was replaced with a new \SI{0.30}{m} one in 2015. Also, the Kodak KAF-16803 CCD based G4-16000 camera was replaced with a new C4-16000 camera equipped with more sensitive GSENSE4040 scientific CMOS sensor in 2020 and a year later with another C3-61000 camera based on Sony IMX455 CMOS sensor, which has even higher quantum efficiency and lower read noise. Image reduction and photometry was performed using freely-available software package \emph{Simple Image Processing System} (SIPS). SIPS has become widely used by citizen scientists to process images and photometry of time-variable sources, but its algorithms and features have not yet been described in the literature. In Appendix~\ref{app:sips}, we address this deficiency and describe some of the capabilities of SIPS.

In total, we have accumulated  $12\,257$ photometric measurements in $99$ observing nights covering years 2012 to 2022. The median number of measurements in one night is $96$, but it has increased to more than $200$ after the technical upgrades. The reason is that larger telescope and more sensitive detectors allowed reduction of the exposure time and thus increase of  the image cadence. We also adjusted the observing program by starting to measure CzeV343 earlier each season to better utilize long winter nights when CzeV343 was above the horizon. The median photometric uncertainty is $0.006$\,mag. We correct magnitude offsets between individual nights in our global photometric model described in Section~\ref{sec:global}. In Figure~\ref{fig:lc_comp}, we show three of our ground-based light curves obtained on consecutive nights. Appendix~\ref{app:ground} provides a  list of our photometric measurements (Table~\ref{tab:ground}) and plots of all our light curves (Figs.~\ref{fig:ground_lc1} and \ref{fig:ground_lc2}). 

\subsection{TESS 2-minute photometry}
\label{obs:tess}

TESS observed CzeV343 with 2-min cadence in Sectors 43, 44, and 45 based on Cycle 4 program G04105 (PI Pejcha). We used {\tt Lightkurve} package \citep{lightkurve18} to download the Target Pixel Files and to extract the aperture photometry using the default mask. Observations in each TESS sector have a break in the middle and as a result the data are naturally divided into six segments. To mask data affected by instrumental effects, we removed points taken within $0.75$\,days of the approximate midpoints between segments. We also remove the first and last $0.5$\,days of observations. To facilitate joint fitting with ground-based data, we convert the fluxes and uncertainties to magnitudes. In total, we retain 43 453 data points. The median photometric uncertainty is $0.009$\,mag. We find that magnitude offsets of individual segments are the only required additional correction and we fit the magnitude offsets together with our DEB model presented in Section~\ref{sec:global}. In Figure~\ref{fig:lc_comp}, we show an excerpt from our 2-min light curve along with simultaneously taken ground-based data. All light curves are shown in Appendix~\ref{app:tess} (Figs.~\ref{fig:tess_lc1} and \ref{fig:tess_lc2}).

\subsection{TESS full-frame images}
\label{obs:ffi}

TESS also recorded CzeV343 on full-frame images (FFI) in Sector 19. Since we took very few ground-based data in that observing season, we decided to include the FFI observations despite the stronger instrumental signals affecting the photometry. We used {\tt Lightkurve} to download $20\times20$ pixel cutouts from FFIs, and performed aperture photometry  with a $3\times 3$ pixel square aperture centered on CzeV343. We masked data at the beginning, the end, and around the middle of the sector with the same algorithm as for the 2-min data. We retain a total of 1006 measurements. To correct for instrumental effects, we selected $196$ background pixels using the threshold method and found the first $15$ orthogonal principal vectors, $\text{PCA}_j$. Since the amplitude and timescale of CzeV343 variability are similar to instrumental effects, we cannot simply fit and subtract the PCA vectors. Instead, we fit the coefficients together with our binary model on the flux level as explained in Section~\ref{sec:global}. We ignore the smearing of the light curves introduced by the FFI 30-min exposure time. Our FFI light curves corrected for instrumental effects are shown in Appendix~\ref{app:tess} (Fig.~\ref{fig:ffi}).

\subsection{Spectra}
\label{obs:spectra}

\begin{figure}
    \centering
    \includegraphics[width=0.48\textwidth]{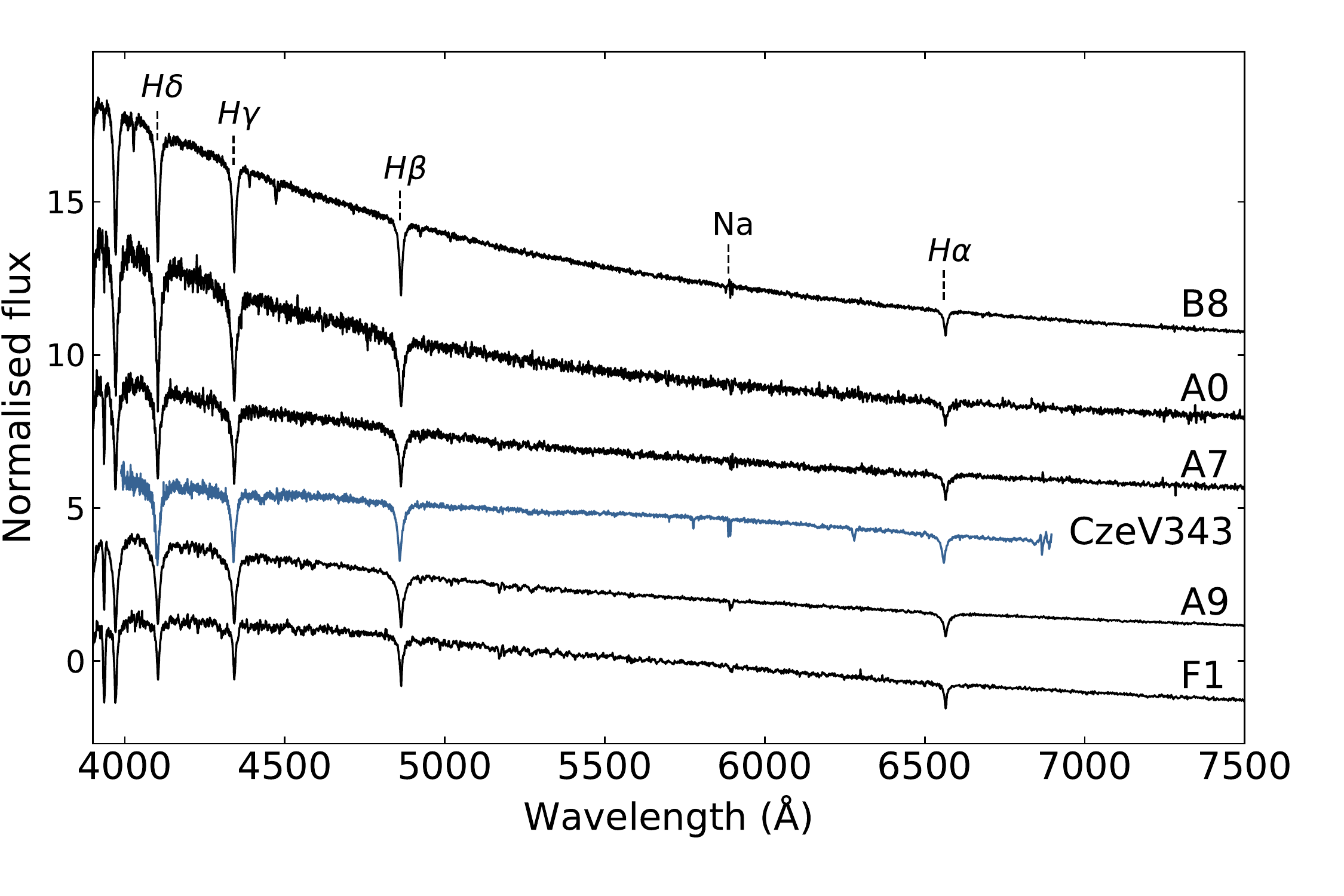}
    \caption{OSMOS spectrum of CzeV343 along with several solar-metallicity spectral templates from PyHammer \citep{pyhammer1,pyhammer2}. For clarity of the display, the spectra were rescaled and shifted.}
    \label{fig:osmos}
\end{figure}

\begin{figure}
    \centering
    \includegraphics[width=0.48\textwidth]{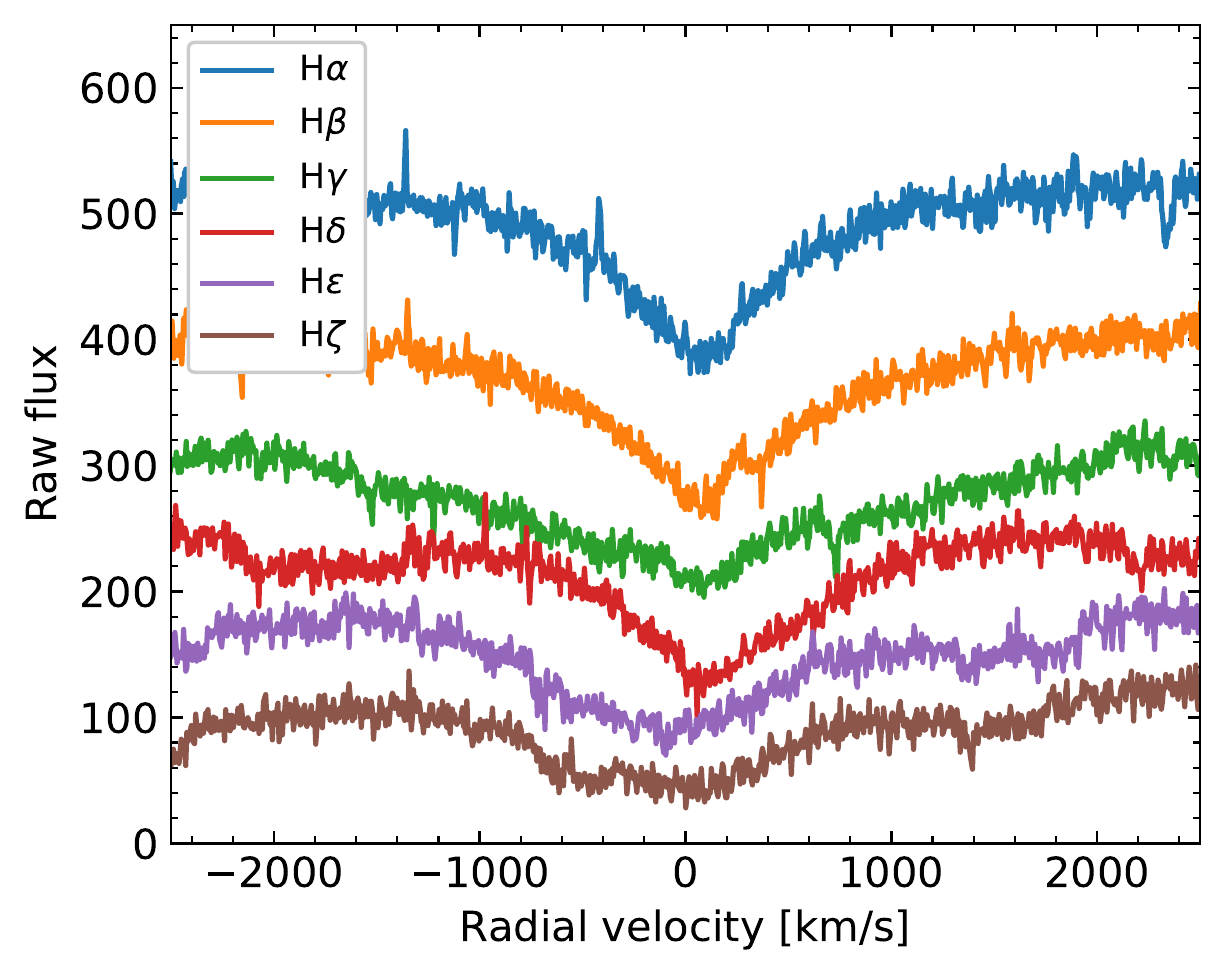}
    \caption{Line profiles of Balmer series lines from our APO spectrum of  CzeV343. The individual profiles were shifted vertically for clarity.}
    \label{fig:apo}
\end{figure}

On October 7 2012 12:07 UTC we obtained a single 10-minute spectrum with the OSMOS spectrograph on the 2.4m Hiltner telescope at MDM Observatory. Unfortunately, spectra of wavelength calibration lamps were not obtained on this night due to technical problems. We used the calibration information available on the instrument website\footnote{\url{https://www.astronomy.ohio-state.edu/MDM/OSMOS/wave/index.html}} to get a crude wavelength correction. We also obtained a rough flux calibration by comparing to the spectrum of Feige 110 obtained during the same observing run. We manually removed artifacts due to cosmics and dead CCD columns.  We show the resulting spectrum in Figure~\ref{fig:osmos}. The broad absorption lines are compatible with an early-type star.

On January 29 2013 07:21 UTC we obtained a single 20-minute exposure of CzeV343 with the echelle spectrograph on the \SI{3.5}{m} telescope at the Apache Point Observatory. The spectrum was processed in a standard way in IRAF echelle package. The individual echelle orders were interpolated on a common wavelength scale and the flux from overlapping orders was co-added. The spectrum is similar to our OSMOS spectrum in the sense that it shows broad Balmer absorption lines. In Figure~\ref{fig:apo}, we show profiles of Balmer series lines. Unfortunately, the lines are often spread over multiple echelle orders, which leads to instrumental waves superimposed on the line profiles and precludes proper disentanglement of the individual components.

\section{Analysis}
\label{sec:analysis}

In this Section, we describe our analysis of photometric measurements starting with the classical $O-C$ analysis (Sect.~\ref{sec:classical}), which motivates the full photometric model (Sect.~\ref{sec:global}). We also present the spectral classification (Sect.~\ref{sec:spectral}) and stellar properties from Gaia DR3 (Sect.~\ref{sec:gaia}).

\subsection{O-C analysis of photometry}
\label{sec:classical}

\begin{figure}
    \centering
    \includegraphics[width=0.48\textwidth]{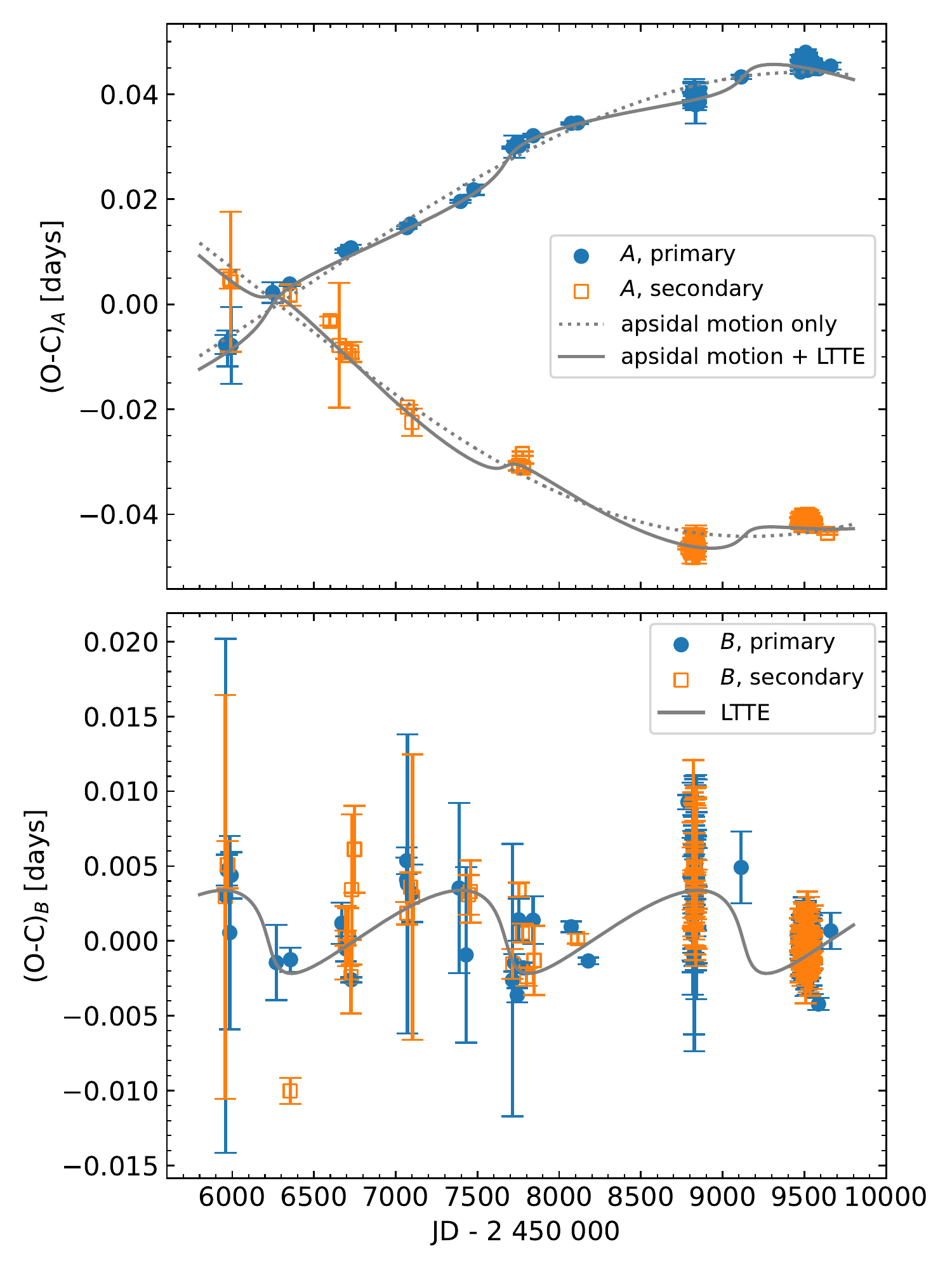}
    \caption{$O-C$ diagram of binary $A$ (top panel) and binary $B$ (bottom panel) calculated by determining minima timings in all of our photometric data. Primary eclipses are shown in blue while secondary eclipses are shown in orange. Gray lines show $O-C$ variations predicted by our best-fit global photometric model from Sect.~\ref{sec:global}. LTTE variations are calculated with Eqs.~(\ref{eq:delays}--\ref{eq:gamma}) and apsidal motion effect is evaluated using expressions from \citet{gimenez83}.}    \label{fig:oc_orig}
\end{figure}

We use the freely available \emph{SImple LIght CUrve Processing System} (SILICUPS) described in Appendix~\ref{app:sips} to determine minima timings of both binaries in CzeV343. The timings are obtained by fitting analytical functions to light curve segments. In Figure~\ref{fig:oc_orig}, we show the resulting $O-C$ diagram with respect to best-fit ephemeris with constant period. We see that in binary $A$ the primary and secondary minima are diverging, which is a clear sign of apsidal motion. In the last three seasons we can recognize that the trends of primary and secondary minima are starting to turn back together, which implies that the apsidal motion period is around $30$ to $40$ years. In addition, we see that there are wiggles occuring in phase for both primary and secondary minima. Binary $B$ has circular orbit and does not show signs of apsidal motion. Instead, we see an oscillating pattern with a timescale of about $1500$\,days and an amplitude of about $0.002$\,days, which has opposite phase than the wiggles seen in binary $A$.

It is likely that the wiggles seen in both $A$ and $B$ are related to the mutual orbit, but it is not immediately clear whether the signal is due to light-travel time effect (LTTE) delay (Roemer effect) or due to dynamic perturbation to the orbits of the inner binaries. According to \citet{borkovits15}, the amplitude of LTTE of binary $A$ is
\begin{multline}
    \Delta_{\text{LTTE},A} \approx 0.0118\,\text{d}\ \left(\frac{P_{AB}}{1454\,\text{d}}\right)^{2/3} \left(\frac{M_B}{2.6\,\msun}\right) \left(\frac{M_{AB}}{5.6\,\msun}\right)^{2/3} \times \\ \times \sin i_{AB} \sqrt{1-e_{AB}^2\cos^2\omega_{AB}},
    \label{eq:ltte}
\end{multline}
where $P_{AB}$ is the mutual orbital period, $M_A$ and $M_B$ total masses of binaries $A$ and $B$,  $M_{AB} = M_A+M_B$ is the total mass of the entire quadruple, $e_{AB}$ is the eccentricity of mutual orbit, $\omega_{AB}$ is the argument of periastron, and $i_{AB}$ is the inclination of the mutual orbit with respect to the observer. \citet{borkovits15} also provided the amplitude of dynamical perturbations on the timescale of $P_{AB}$ to binary $A$ as
\begin{multline}
    \Delta_{\text{dyn},A} \approx 1.37\times 10^{-4}\,\text{d}\ \left(\frac{P_A}{1.2\,\text{d}}\right)^2 \left(\frac{P_{AB}}{1454\,\text{d}}  \right)^{-1} \left( \frac{M_B}{2.6\,\msun} \right) \left(\frac{M_{AB}}{5.6\,\msun}\right)^{-1} \times \\ \times
    (1-e_A^2)^{1/2} (1-e_{AB}^2)^{-3/2},
    \label{eq:dyn}
\end{multline}
where $e_A$ is the eccentricity of binary $A$. 

Comparing these expressions to the amplitude seen in Figure~\ref{fig:oc_orig}, we see that dynamical perturbations are too small to explain the observed variations. The signal comes from LTTE, but the observed amplitude is significantly smaller than the maximum value. This implies that the mutual orbit is seen quite face-on. We quantify this finding and discuss its implications in Section~\ref{sec:mutual}.

\subsection{Global model of photometry}
\label{sec:global}

\renewcommand{\arraystretch}{1.2}
\begin{table*}
\caption{Double eclipsing binary light curve fit results. We give name of the parameter, the range of the uniform prior, and median value of the posterior together with confidence interval encompassing $95.4\%$ of the probability.}              
\label{tab:fit}      
\centering                                      
\begin{tabular}{l c c }          
\hline\hline                        
Parameter name & Allowed range & Value \\    
\hline                                   
\multicolumn{3}{c}{Global parameters}\\
\hline
ratio of fluxes (ground) & $10^{-5}\le \beta_{\rm ground} \le 1.0$ & $0.4495^{+0.0194}_{-0.0168}$\\
ratio of fluxes (TESS) & $10^{-5}\le \beta_{\rm TESS} \le 1.0$ & $0.4513^{+0.0194}_{-0.0168}$\\
additional light (ground) & $10^{-5}\le F_{5,\rm ground} \le 1.0$ & $0.1970^{+0.0244}_{-0.0297}$\\
additional light (TESS) & $10^{-5}\le F_{5,\rm TESS} \le 1.0$ & $0.2811^{+0.0220}_{-0.0266}$\\
\hline
\multicolumn{3}{c}{Binary $A$}\\
\hline
Epoch of primary conjunction (JD - 2 450 000) & $6246.9\le T_{0} \le 6247.9$ & $6247.40006^{+0.00029}_{-0.00031}$\\
Orbital period (days) & $1.2092\le P \le 1.2094$ & $1.209340820^{+0.000000081}_{-0.000000080}$\\
Central surface brightness ratio & $0.1\le \Theta \le 2.0$ & $0.6609^{+0.0022}_{-0.0022}$\\
Sum of radii (in units of semi-major axis)& $0.0\le r_1+r_2 \le 1.0$ & $0.43474^{+0.00098}_{-0.00098}$\\
Ratio of radii & $0.0\le r_1/r_2 \le 5.0$ & $0.7208^{+0.0062}_{-0.0062}$\\
Cosine of inclination & $0.0\le \cos i \le 1.0$ & $0.0444^{+0.0056}_{-0.0060}$\\
Mass ratio & $0.0\le q \le 1.5$ & $0.966^{+0.039}_{-0.039}$\\
Eccentricity times cosine of argument of periastron & $-1.0\le e\cos\omega \le 1.0$ & $0.00192^{+0.00050}_{-0.00050}$\\
Eccentricity times sine of argument of periastron & $-1.0\le e\sin\omega \le 1.0$ & $0.11432^{+0.00019}_{-0.00019}$\\
Apsidal precession rate (rad/day) & $-0.001\le d\omega/dt \le 0.001$ & $0.0005149^{+0.0000031}_{-0.0000031}$\\
\hline
\multicolumn{3}{c}{Binary $B$}\\
\hline
Epoch of primary conjunction (JD - 2 450 000) & $5984.1\le T_{0} \le 5984.9$ & $5984.47189^{+0.00040}_{-0.00038}$\\
Orbital period (days) & $0.8068\le P \le 0.8070$ & $0.806871019^{+0.000000077}_{-0.000000076}$\\
Central surface brightness ratio & $0.1\le \Theta \le 2.0$ & $1.0299^{+0.0216}_{-0.0209}$\\
Sum of radii (in units of semi-major axis) & $0.0\le r_1+r_2 \le 1.0$ & $0.56322^{+0.00561}_{-0.00555}$\\
Ratio of radii & $0.0\le r_1/r_2 \le 5.0$ & $1.1649^{+0.1870}_{-0.1616}$\\
Cosine of inclination & $0.0\le \cos i \le 1.0$ & $0.2691^{+0.0198}_{-0.0198}$\\
Mass ratio & $0.0\le q \le 1.2$ & $0.5712^{+0.2011}_{-0.1730}$\\
Eccentricity times cosine of argument of periastron & $-1.0\le e\cos\omega \le 1.0$ & $0.00050^{+0.00031}_{-0.00031}$\\
Eccentricity times sine of argument of periastron & $-1.0\le e\sin\omega \le 1.0$ & $-0.00162^{+0.00213}_{-0.00207}$\\
\hline
\multicolumn{3}{c}{Mutual orbit}\\
\hline
Epoch of primary conjunction (JD - 2 450 000) & $2800.0\le T_{0} \le 7800.0$ & $6224.8^{+20.7}_{-25.6}$\\
Orbital period (days) & $500.0000\le P \le 2000.0000$ & $1453.9^{+10.2}_{-8.4}$\\
Eccentricity & $0.0\le e \le 1.0$ & $0.703^{+0.091}_{-0.073}$\\
Argument of periastron (rad) & $0.0\le \omega \le 2\pi$ & $0.225^{+0.105}_{-0.104}$\\
Modulation amplitude of $A$ (days) & $0.00\le D_A \le 0.01$ & $0.001754^{+0.000118}_{-0.000109}$\\
Modulation amplitude of $B$ (days) & $-0.01\le D_B \le 0.00$ & $-0.002196^{+0.000150}_{-0.000163}$\\
\hline
\end{tabular}
\end{table*}

Light curves of the two eclipsing binaries in a DEB overlap, which leads to complex features and complicates interpretation of the observations. In DEBs with near-resonant periods, the alignment of the features slowly drifts, which can bias minima timings obtained in a traditional way. For example, when binary $A$ slowly brightens (for example due to ellipsoidal variations) over the course of the eclipse of binary $B$ then fitting of minima of $B$ will be biased if the variation from the other binary is not simultaneously accounted for. Effects like these might contribute to relatively high scatter seen in the Figure~\ref{fig:oc_orig}.

To properly characterize properties of CzeV343, we extend the analysis presented by \citet{cagas12} to include apsidal motion, mutual orbit, and simultaneous fitting of all types of photometric data. In our model, the magnitude at time $t_i$ is
\begin{equation}
    m(t_i) = -2.5\log_{10}F_\text{tot}(t_i) + \sum_j c_j \Delta_{ij},
    \label{eq:mag}
\end{equation}
where $F_\text{tot}$ is the total flux from the system, $c_j$ are coefficients describing magnitude offsets of each segment (either one ground observing night or one half of TESS sector), and $\Delta_{ij}$ is unity for $t_i$ within segment $j$ and zero otherwise. The total flux is
\begin{multline}
F_\text{tot}(t_i) = (1-F_5)(1-\beta)F_A(t'_i) +\\ + (1-F_5)\beta F_B(t''_i) + F_5 + \delta_\text{FFI}\sum_j b_j \text{PCA}_j(t_i),
\label{eq:ftot}
\end{multline}
where $F_A$ and $F_B$ are fluxes of binaries $A$ and $B$, $\beta$ parameterizes relative flux of $A$ and $B$, $F_5$ accounts for any additional light such as due to an additional unresolved component or blending with nearby stars, and $b_j$ are coefficients that project the residuals of the binary model to the principal component vectors $\text{PCA}_j$ in TESS FFI data, which is facilitated by setting $\delta_\text{FFI}$ to unity for TESS FFI data and to zero otherwise. The model is constructed to ensure that the normalization of the total flux does not depend on the actual values of $\beta$ and $F_5$.

To take into account LTTE, we evaluate the binary fluxes at shifted times $t'_i$ and $t''_i$, which are defined as
\begin{subequations}
\begin{eqnarray}
t'_i &=& t_i - D_A \gamma(t_i),\\
t''_i &=& t_i - D_B \gamma(t_i),
\end{eqnarray}
\label{eq:delays}
\end{subequations}
where $D_B \le 0 \le D_A$ are amplitudes of the LTTE variation of the two binaries that depend on $M_A$, $M_B$, and $i_{AB}$. Function $\gamma(t)$ was calculated by \citet{irwin52} as
\begin{equation}
    \gamma(t) = \frac{1-e_{AB}^2}{\sqrt{1 - e_{AB}^2\cos^2\omega_{AB}}} \frac{\sin(\nu_{AB}+\omega_{AB})}{1+e_{AB}\cos\nu_{AB}},
    \label{eq:gamma}
\end{equation}
where $\nu_{AB}$ is the true anomaly, which is related to $P_{AB}$ by the Kepler equation. We omitted the last term in the \citet{irwin52} expression, because it only introduces constant shift in time. We use Kepler equation solver\footnote{\url{https://github.com/dfm/kepler.py}} extracted from the package {\tt exoplanet} \citep{exoplanet:joss}.

To calculate $F_A$ and $F_B$ we use the eclipsing binary model {\tt eb}\footnote{\url{https://github.com/mdwarfgeek/eb}} \citep{irwin11}, which is based on JKTEBOP \citep{etzel81,southworth04,southworth07,southworth09,southworth10}. Each binary model is parameterized by the epoch of primary conjuction $T_0$, orbital period $P$, central surface brightness ratio $\Theta$, sum $r_1+r_2$ and ratio $r_1/r_2$ of relative radii, cosine of inclination $\cos i$, photometric mass ratio $q$, and two eccentricity parameters $e\cos\omega$ and $e\sin\omega$. For binary $A$ we also include the apsidal precession rate $d\omega/dt$. We use standard values appropriate for early type stars of limb darkening ($u_1 = 0.05$, $u_2 = 0.6$), gravity darkening ($0.25$), and albedo ($0.4$). 

Our metric for finding the best fit is
\begin{equation}
    \chi^2 = \sum_i \left(\frac{m_i - m(t_i)}{\sigma_i}\right)^2,
\end{equation}
where $m_i$ and $\sigma_i$ are observed magnitudes and their uncertainties at time $t_i$. Because the ground-based and TESS data have different spectral response, we fit separate $\beta$  and $F_5$ for each of these datasets. Similarly, we should also fit separately the surface brightness ratio $\Theta$, but we find that in such case $q$ and $r_1/r_2$ become unconstrained for ground-based data. Instead, we find that using joint $\Theta$ leads to more stable fits that still sufficiently well model both datasets. In total, our model has $152$ free parameters.

We find the minimum of $\chi^2$ using the routine {\tt least\_squares} in {\tt scipy}. Our best-fit model has $\chi^2 = 63\,395$ for a total of 56\,564 degrees of freedom. The contributions of individual datasets to the joint $\chi^2$ are: ground-based $\chi^2 = 12\,838$ for 12\,257 points, TESS 2-min $\chi^2 = 49\,681$ for 43\,453 points, and TESS FFI $\chi^2 = 1028$ for $1006$ points. These numbers indicate that none of the datasets has its uncertainties under- or overestimated and that each dataset contributes to the final $\chi^2$ proportionally to its size. We illustrate the best-fit model of the light curves in Figure~\ref{fig:lc_comp} and we visualize all fits in Appendices~\ref{app:ground} and \ref{app:tess}.

With the best-fit values as a starting point, we explore the distribution of likelihood $\ln \mathcal{L} = -\chi^2/2$ in the space of parameters with the Markov Chain Monte Carlo (MCMC) package {\tt emcee} \citep{emcee}. We find that it is not feasible to directly explore the full $152$-dimensional space due to the memory requirements of {\tt emcee}. However, majority of the free parameters are coefficients $c_j$, which are simply offsets of individual segments. We modify the likelihood function so that $c_j$ are determined simply as a weighted-mean offset of each segment from the model described by the remaining parameters. This reduces the dimension for MCMC exploration to $30$ without fundamental loss of information, which we verified with shorter chains run on the full parameter space. With our modified likelihood, we set uniform priors for all parameters and run $90$ chains for $500\,000$ steps each. The autocorrelation length of the chains is about $1000$ steps. Projected distributions of the posterior are shown in Appendix~\ref{app:posteriors}. 

In Table~\ref{tab:fit}, we report the results of our model and its uncertainties. We report the median value of each parameter from the Markov chains after discarding the initial $10\,000$ steps and thinning by $1000$. The confidence intervals capture $95.4\%$ of the probability. In the rest of the paper, we quote the same confidence interval for quantities based on our photometric model unless stated otherwise. Our results are broadly consistent with values previously reported by \citet{cagas12} with one exception. \citet{cagas12} reported $e_A = 0.18$, but we find $e_A=0.11$. This can be explained by the limited timespan of the earlier work, when the phase of the apsidal motion was very close to $0.5$ and the eccentricity was constrained only by the eclipse duration rather than eclipse timing. Here, we have covered a substantial fraction of the apsidal motion cycle, which gives us a more robust determination of $e_A$. We caution that parameters like $\Theta$, $q$, and $r_1/r_2$ are partially degenerate, especially in the case of binary $B$, and the values we obtained are likely not robust. Nonetheless, this is not a significant problem for us, because we are primarily interested in the timing properties, where these parameters do not play a role. We will discuss physical parameters of the binaries in Sect.~\ref{sec:physical}. 

We note that the uncertainties on quantities like $d\omega_A/dt$ and $P_{AB}$ are relatively small, especially considering that the timespan of our observations covers about two cycles of $P_{AB}$ and about a third of the apsidal motion cycle of binary $A$. One possible explanation is that the MCMC analysis is not sufficiently converged, but we verified that the length of our Markov chains sufficiently exceeds their autocorrelation scale. Another possible explanation is that in our global model timing of all individual measurements enters the solution, whereas classical $O-C$ analysis is performed on a much smaller number of derived quantities with potentially additional noise introduced by the intermediate layer of minima fitting. Alternatively, there might be red noise that is not fully taken into account in our model as was found in a similar analysis by \citet{torres17}.

Finally, both \citet{cagas12} and us find substantial amount of constant additional light in CzeV343. This could be caused by inadequacy of the binary model, which does not appropriately reproduce the observations especially around minima. Alternatively, there is an additional component in CzeV343, which would make it a quintuple star similar to V994~Her \citep{zasche16}. However, \citet{cagas12} did not find any photocenter variations, which means that the angular distance of the hypothetical additional component must be small. The higher value of $F_5$ in TESS data can be explained either by red color of the fifth star or as additional flux from nearby stars. In any case, this finding is not important for our subsequent analysis and we will therefore not discuss $F_5$ again.

\subsection{Spectra}
\label{sec:spectral}

We use the package {\tt PyHammer} \citep{pyhammer1,pyhammer2} to match our OSMOS spectrum to a library of spectral templates. The best match is to an A7 template at metallicity $\text{[Fe/H]}=-1.5$. We consider such low metallicity unlikely\footnote{Alternatively, low metallicity could be explained by the $\lambda$~Boo phenomenon in A-type stars, which manifests by depletion of iron but relatively normal abundances of many other elements.}, although CzeV343 is located $3.7_{-0.2}^{+0.4}$\,kpc away almost exactly opposite to the Milky Way center ($b=1.5^\circ$, $l=178^\circ$) \citep{gaiaedr3dist}. {\tt PyHammer} does not allow constraints on metallicity during fits, but in Figure~\ref{fig:osmos} we compare our spectrum to several solar-metallicity templates. We see that still our spectrum best resembles late-A stars. 

Our APO spectrum is unfortunately not useful for spectral classification, because it is not flux-calibrated and the broad absorption lines often cover multiple Echelle orders, which leads to artificial wiggles. However, the high resolution of the spectrum allows us to get an independent estimate of the reddening from the equivalent width of Na I D lines. We used \citet{poznanski12} relations for both sodium lines and combined their uncertainties with the estimates of equivalent width uncertainty. We obtain $E(B-V) = 0.22^{+0.08}_{-0.06}$\,mag. For $R_V = 3.1$ extinction law of \citet{cardelli89}, this corresponds to extinction at 550\,nm of $0.68^{+0.25}_{-0.19}$\,mag and in the Gaia $G$ band of approximately $0.61^{+0.22}_{-0.17}$\,mag.

\subsection{Gaia DR3}
\label{sec:gaia}

CzeV343 is included in the recently released Gaia DR3 catalog, which provides  astrometry and stellar parameters based on low-resolution BP and RP spectra and other observations and models \citep{gaiadr3,apsis}.  The observed $G$-band mean magnitude is $13.54$\,mag and we verified with the Gaia epoch photometry that this value is representative of magnitudes near the maximum brightness.  The inverse parallax distance to CzeV343 is $4.5 \pm 0.5$\,kpc. This estimate does not take into account Lutz--Kelker and similar biases. After taking these effects into account, \citet{gaiaedr3dist} provided distances based on Gaia EDR3 parallaxes. For CzeV343 they give $3.68^{+0.37}_{-0.26}$\,kpc, which is noticeably closer than the inverse parallax distance.

Since CzeV343 is composed of four eclipsing stars, the interpretation of tabulated stellar parameters  in Gaia DR3 is not straightforward. As we will show in Section~\ref{sec:physical}, the important parameter is the effective temperature of the most luminous star in the system. Gaia DR3 provides stellar parameters determined in several ways. GSP-Phot combines BP and RP spectra with photometry, astrometry, and stellar models. For CzeV343, the best-fit model is from A star templates, which gives $T_\text{eff} = 10793^{+62}_{-44}$\,K, $\feh =-1.25^{+0.11}_{-0.26}$,  $G$-band extinction $A_G = 1.233^{+0.003}_{-0.003}$\,mag, and distance $2.23^{+0.05}_{-0.06}$\,kpc. GSP-Phot using MARCS models gives $T_\text{eff} = 7168^{+596}_{-212}$\,K, $\feh =-0.42^{+0.01}_{-0.01}$, $A_G = 0.38^{+0.25}_{-0.10}$\,mag, and distance $3.34^{+0.27}_{-0.50}$\,kpc. A separate pipeline MSC models the spectra as a combination of two components with different temperatures and fluxes. The brighter component has $T_\text{eff} = 7184^{+685}_{-1522}$\,K, the system metallicity is $\feh =0.11^{+0.27}_{-0.35}$,  the extinction is $A_G = 0.52^{+0.21}_{-0.52}$\,mag, and the distance is $2.4^{+1.6}_{-1.7}$\,kpc. The range of inferred parameters is relatively wide and we discuss the implications in Sect.~\ref{sec:physical}.

 As we will show in Sect.~\ref{sec:mutual}, the two binaries are on a mutual orbit with semi-major axis of about $4.5$\,AU, which translates to angular separation of about $1.5$\,mas at a distance of $3$\,kpc. Although CzeV343 is not included in the catalog of non-single stars in Gaia DR3, the astrometric solution exhibits excess noise of $0.122$\,mas with a significance of about $12$. Following \citet{hwang20}, the expected astrometric noise for an equal-flux pair of sources exhibiting 25\% photometric variability similar to CzeV343 is about $0.37$\,mas, which is similar to the observed Gaia value. It is possible that Gaia astrometry is affected by photocentroid shifts caused by partially resolving the mutual orbit.

\section{Properties of CzeV343}
\label{sec:results}

In this Section, we explore properties of CzeV343 based on our fits of light curves and spectra. We begin by discussing basic physical parameters (Sect.~\ref{sec:physical}) followed by the apsidal motion in binary $A$ (Sect.~\ref{sec:apsidal}), spin state of the stars (Sect.~\ref{sec:rotation}), mutual orbit (Sect.~\ref{sec:mutual}), effects of nodal precession (Sect.~\ref{sec:nodal}), and resonance of the orbital periods (Sect.~\ref{sec:resonance}).

\begin{figure*}
    \centering
    \includegraphics[width=\textwidth]{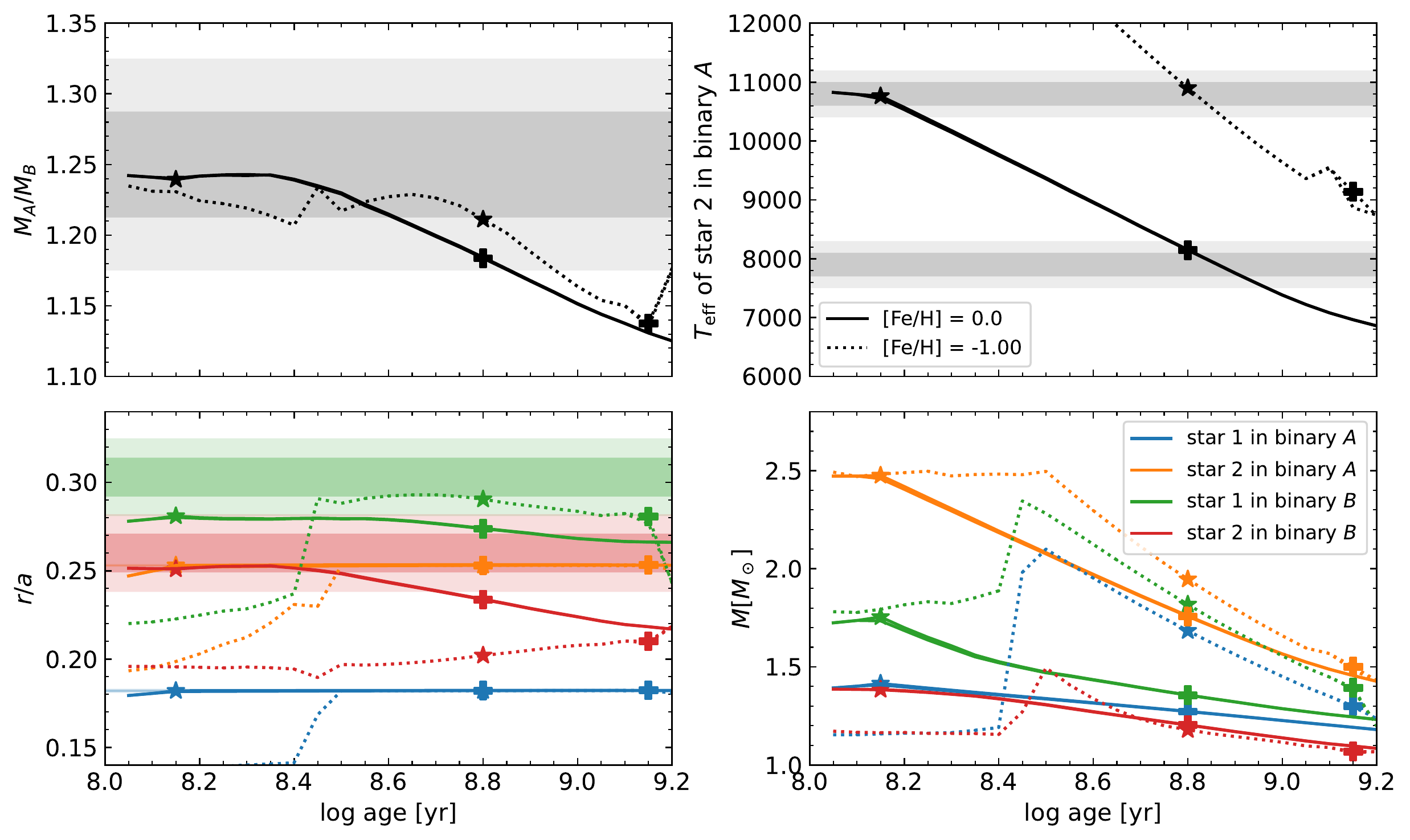}
    \caption{Constraints on physical properties of stars in CzeV343 using MIST isochrones. We show three observables that we are trying to simultaneously fit: mass ratio of the two binaries constrained by LTTE (top left), effective temperature of star 2 in binary $A$ (top right), and the four relative radii of individual components from our photometric model (bottom left). In these three panels, lighter and darker bands show 1 and 2$\sigma$ confidence intervals of the constrained parameters. In the top right diagram, we show two gray bands corresponding to two choices of $T_\text{eff,prior}$. Bottom right panel shows the best-fit initial stellar masses. In all panels, lines show best-fit models as a function of isochrone age for two metallicities as explained in the legend. The best-fit model for each metallicity is marked separately for $T_\text{eff,prior} = 7900$\,K (plus sign) and $T_\text{eff,prior} = 10800$\,K (star).}
    \label{fig:iso}
\end{figure*}

\subsection{Physical properties}
\label{sec:physical}

We aim to derive radii and masses of all four stars in CzeV343. Our main observables are relative radii from our photometric model: $r_{A,1} = 0.1821 \pm 0.0006$, $r_{A,2} = 0.2526 \pm 0.0004$, $r_{B,1} = 0.303 \pm 0.010$, and $r_{B,2} = 0.260 \pm 0.011$.  Notice that uncertainties on radii in binary $B$ are more than an order of magnitude larger than in binary $A$. Another constraint comes from the amplitudes of LTTE, which directly constrain the mass ratio of the two binaries as $M_A/M_B = |D_B/D_A| = 1.252 \pm 0.038$. In this section, the reported uncertainties from MCMC of our photometric model are $1\sigma$ confidence intervals. So far all of the observables are relative quantities so we require something else to obtain absolute masses and radii. There are two options: spectral types from ours OSMOS spectrum or effective temperature from Gaia DR3. 

First, we discuss our OSMOS spectrum. Our photometric model puts the time of the exposure to the moment when binary $A$ was just entering the secondary eclipse and binary $B$ was just leaving the eclipse. This implies that the spectrum is dominated by the flux of binary $A$. In addition, we also found that $\beta < 0.5$, which makes binary $A$ the dominant component even outside of the eclipse. Furthermore, our photometric model implies that  star $2$ is the brighter and larger star in binary $A$. Taken together, this implies that our spectra should be dominated by star $2$ in binary $A$. We therefore put a constraint on the effective temperature of this star to $T_\text{eff,prior} = 7900 \pm 200$\,K based on empirical relation between spectral types and effective temperature of \citet{pecaut12} and \citet{pecaut13}.

Second, Gaia DR3 provides two distinct values for effective temperature. One is compatible with the estimate from our OSMOS spectrum and the second gives much higher value of about 10800\,K corresponding to late B or early A type spectrum. Since the Gaia result is an average over many epochs, we can assume that the value roughly corresponds to the brightest star in the system, which is star $2$ in binary $A$. We therefore investigate a second possibility that $T_\text{eff,prior} = 10800 \pm 200$\,K.

In total, we have six constraints for eight parameters, which implies that without additional information we cannot constrain all radii and masses independently. If we assume that all four components have the same age, metallicity, and have not experienced any kind of binary interaction yet, then we can use theoretical stellar isochrones to fill in the missing information. We use MIST isochrones \citep{dotter16,choi16} to calculate instantaneous mass, radius, and effective temperature for a star of given initial mass, age, and metallicity. We combine our observational constraints to a likelihood of the form
\begin{multline}
    \ln \mathcal{L} = \frac{1}{2}\sum_i \left(\frac{R_i/a - r_i}{\sigma_i}\right)^2 + \left(\frac{T_{\text{eff},A2} - T_\text{eff,prior}}{200\,\text{K}}  \right)^2 + \\ + \left(\frac{M_A/M_B - 1.25}{0.0375} \right)^2,
    \label{eq:likeli}
\end{multline}
where $\sigma_i$ are uncertainties, and the factor $1/2$ in front of the first term gives relatively less weight on relative radii. By minimizing $\ln\mathcal{L}$, we find best-fit values of the initial masses of all four components. We choose to perform the fit separately for each age and metallicity and for two different values of $T_\text{eff,prior}$ to better understand how our model behaves.

\begin{table}[]
    \caption{Physical parameters of stars in CzeV343 from solar-metallicity isochrones.}
    \label{tab:my_label}
    \centering
    \begin{tabular}{c cc c }
    \hline
    \hline
    Star & Mass $[\msun]$ & Radius $[\rsun]$ & $T_\text{eff}$ [K]\\
    \hline
    \multicolumn{4}{c}{Fiducial: $T_\text{eff,prior} = 7900 \pm 200$\,K, age $\approx 630$\,Myr}\\
    \hline
         Binary $A$, star 1& $1.27$ & $1.26$ & $6570$\\
         Binary $A$, star 2& $1.76$ & $1.75$ & $8140$\\
         Binary $B$, star 1& $1.36$ & $1.37$ & $6820$\\
         Binary $B$, star 2& $1.20$ & $1.17$ & $6370$\\
    \hline
        \multicolumn{4}{c}{$T_\text{eff,prior} = 10800 \pm 200$\,K, age $\approx 140$\,Myr}\\
    \hline
         Binary $A$, star 1& $1.42$ & $1.37$ & $7000$\\
         Binary $A$, star 2& $2.48$ & $1.90$ & $10760$\\
         Binary $B$, star 1& $1.75$ & $1.50$ & $8470$\\
         Binary $B$, star 2& $1.38$ & $1.34$ & $6900$\\
    \hline
    \end{tabular}
\end{table}

In Figure~\ref{fig:iso}, we show the best-fit masses and radii as a function of stellar age. We can very well reproduce radii in binary $A$ and the temperature of star $2$ in binary $A$ for both $T_\text{eff,prior}$. In most models that we tried we find that the isochrones predict somewhat smaller radii in binary $B$, but here the observed values have much higher uncertainties than in binary $A$. Models with $\feh = -1.00$ lead to very high discrepancy in the relative radius of star 2 of binary $B$. Models with $T_\text{eff,prior} = 10800$\,K give better agreement with $M_A/M_B$. In Table~2, we quote the best-fit solar metallicity results for both choices of $T_\text{eff,prior}$.

Are our estimates consistent with other available information? The ratios of total fluxes of stars $1$ and $2$ in binaries $A$ and $B$ in our photometric model are  $\Theta (r_1/r_2)^2\approx 0.343\pm 0.003$ and $1.40\pm 0.20$. The flux ratio between the two binaries is $(1-\beta)/\beta \approx 1.22\pm 0.04$. The ground-based data are taken with an instrument sensitive between about $400$ and $1000$\,nm, with a peak sensitivity around $600\,$nm. TESS has a nearly uniform sensitivity between $600$ and $1000$\,nm. We use MIST synthetic absolute magnitudes to calculate flux ratios in the TESS (Bessell $R$) band for the best-fitting solar-metallicity model with $T_\text{eff,prior} = 7900$\,K for binaries $A$ and $B$, which gives $0.29$ ($0.26$) and $1.66$ ($1.72$). The flux ratio of the two binaries is $2.15$ ($2.32$). Similar numbers are obtained for the model with $T_\text{eff,prior} = 10800$\,K. These results suggest that our physical parameters are broadly compatible with our photometric model except for the flux ratio of the two binaries.

Another comparison can be obtained from the photometric parallax. We used the MIST isochrones to estimate the total $G$-band absolute magnitude of our two best-fit solar-metallicity models. We find that $0.52$ and $1.41$\,mag for $T_\text{eff,prior} = 10800$ and $7900$\,K, respectively. Using the $G$-band extinction based on equivalent width of sodium lines (Sect.~\ref{sec:spectral}), we find distance moduli of $12.41^{+0.25}_{-0.19}$ or $11.52^{+0.25}_{-0.19}$\,mag for the two respective models with uncertainties dominated by the extinction. This corresponds to distances of $3.0^{+0.4}_{-0.3}$ or $2.0^{+0.3}_{-0.2}$\,kpc. The photometric parallax favors the model with higher $T_\text{eff,prior}$. However, models with $\feh = -1.0$ lead to distance moduli higher by a about $0.30$ and $0.53$\,mag, respectively. Furthermore, the MIST models underpredict stellar radii as compared to our light curve solution, which also implies that the photometric distance might be  too close.

Which of the models should we prefer? Although $T_\text{eff,prior} = 10800$\,K gives an overall better agreement with our constraints and with photometric parallax, this value is supported only by one of the pipelines in Gaia DR3, while the other pipelines give effective temperatures compatible with our OSMOS spectrum. We therefore choose the $T_\text{eff,prior} = 7900$\,K and $\feh = 0.00$ results as our fiducial parameters of CzeV343 for the rest of the paper. This choice is somewhat arbitrary, in fact, we could probably construct similarly plausible solutions for a range of $T_\text{eff,prior}$. The uncertainty on stellar masses is best obtained by comparing results for the two metallicities and for the same age, which implies uncertainty of about $10\%$. However, the inferred values are highly correlated in the sense that mass ratios are better constrained than the absolute values.

To summarize, our estimates suggest that binary $A$ is composed of a late A and late F stars with a total mass about $3\,\msun$ and binary $B$ is composed of mid F and  late F stars with a total mass about $2.6\,\msun$. We emphasize that precise estimates of individual properties of all four stars is not the main goal of this paper. What is important is that likely all four stars are above the Kraft break and have radiative envelopes. We caution that our estimates are contingent on the effective temperature of the brightest star in the system, which we were unable to unambiguously constrain. Clearly, a more robust picture would come from performing spectral disentangling and radial velocity analysis on new spectroscopic data, similarly to what was done for V994~Her and other DEBs \citep[e.g.][]{lee08,kostov21},  or by constructing a detailed spectro-photometric model combining light curves, spectral energy distributions, and astrometry \citep[e.g.][]{borkovits21}.  Throughout the rest of the text we discuss implications of having higher stellar masses wherever it is appropriate.  Fortunately, many quantities depend on mass ratios rather than the total mass and the mass uncertainty is often not the fundamental limitation of our analysis.

\subsection{Spins of individual stars}
\label{sec:rotation}

\begin{figure*}
    \centering
    \includegraphics[width=\textwidth]{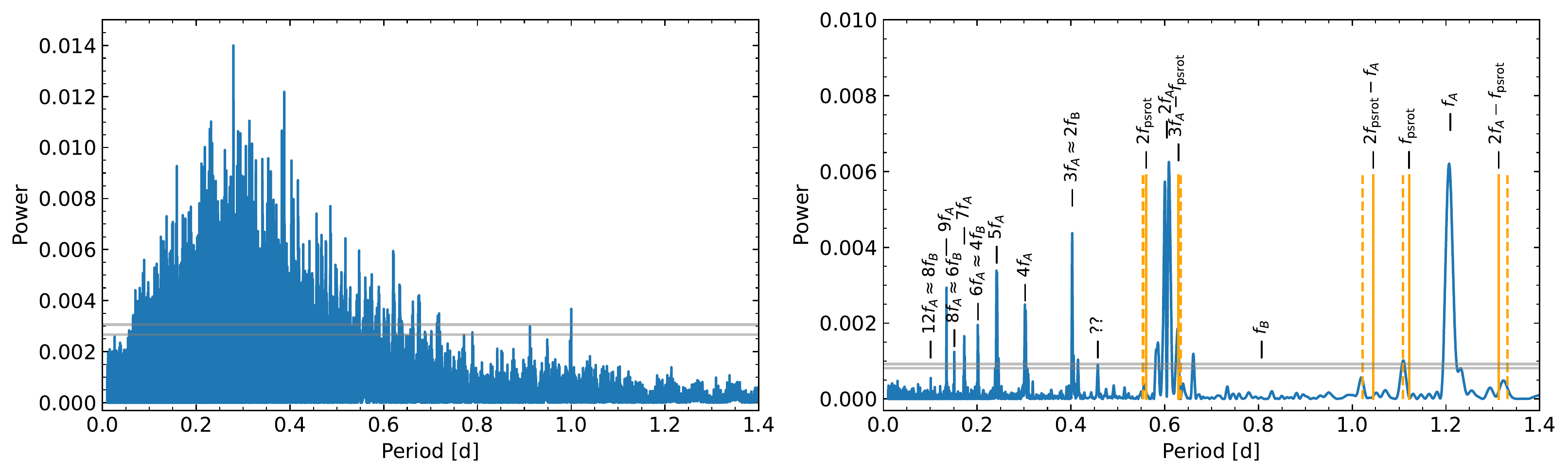}
    \caption{Periodograms of residuals after fitting the double eclipsing binary model. We show separately ground-based data (left) and TESS short-cadence data (right). The two grey horizontal lines in both panels indicate false alarm probability levels of $0.1$ (lower) and $0.01$ (upper). The right panel shows additionally identification of some of the most prominent modes and their harmonics. Here, the frequency is defined as $f = 1/P$. The orange vertical lines in the right panel indicate the pseudo-synchronous rotation frequency and its main harmonics that correspond to the eccentricity found in the best double eclipsing binary model (solid) and $f_\text{psrot}$ increased by $1\%$ relative to Eq.~(\ref{eq:ps}) (dashed).}
    \label{fig:period}
\end{figure*}

The spin state of individual stars in CzeV343 is of interest, because it can constrain history of tidal evolution and can potentially reveal spins misaligned with orbits \citep[e.g.][]{dai18,liang22}. Unfortunately, all stars in CzeV343 seem to have radiative envelopes, which makes presence of prominent stellar spots unlikely as evidenced by the decreasing spot amplitude with stellar effective temperature \citep[e.g.][]{reinhold19,avallone22,david22}. Indeed, we do not see any apparent modulation inconsistent with the our DEB model. 

However, we can probe low-level activity by studying residuals of the DEB model. In Figure~\ref{fig:period}, we show results of Lomb--Scargle periodogram \citep{astropy:2013,astropy:2018} performed separately on ground-based (left panel) and TESS 2-min data (right panel). We can see that despite their long total total time-span, ground-based data are insufficient to reveal any prominent frequencies. Highest powers are reached for periods between $0.2$ and $0.4$\,days, which corresponds to the typical duration of nightly observing runs. 

The situation is different for TESS 2-min data, where we see a number of harmonics associated with the orbital frequency of binary $A$, $f_A = 1/P_A$. These harmonics can be either caused by poor fit due to uncorrected systematics, inadequacy of the underlying binary model based on ellipsoids rather than Roche surfaces, or actual dynamical perturbations to the orbit due to binary $B$ (Eq.~[\ref{eq:dyn}]). We did not find any harmonics of binary $B$ except of the obvious resonances with binary $A$. 

In addition to these modes, we identified a triplet of modes surrounding $f_A$, which does not straightforwardly match any obvious combination of $f_A$ and $f_B$. Instead, we find that these modes approximately correspond to the harmonics of the pseudo-synchronous rotation frequency $f_\text{psrot}$ reached by tidally synchronized stars in an eccentric binary. According to \citet{hut81} and \citet[][p.~161]{eggleton06}, $f_\text{psrot}$ is
\begin{equation}
    \frac{f_\text{psrot}}{f_A} = \frac{1+\frac{15}{2}e_A^2 + \frac{45}{8}e_A^4 + \frac{5}{16}e_A^6}{\left(1-e_A^2\right)^{3/2} \left(1+ 3e_A^2 + \frac{3}{8}e_A^4\right)} \approx 1.079.
    \label{eq:ps}
\end{equation}
In this expression, we neglect the small correction due to spin angular momentum, which is on the level of $10^{-4}$ for binary $A$. The peak at $f_\text{psrot}$ has false alarm probability level smaller than $0.01$, but the other harmonics have higher false alarm probabilities.

Interestingly, the pseudo-synchronous rotation peaks in the periodogram are slightly shifted from what Equation~(\ref{eq:ps}) predicts based on $e_A$ from our photometric model. To bring the results in agreement we would have to increase $f_\text{psrot}$ by about $1\%$. There are four possible explanations. First, Equation~(\ref{eq:ps}) is inaccurate, because it is based on equilibrium tidal theory, which only approximates the real tidal field. Second, our photometric eccentricity is lower by approximately $8\%$ than the true value, which we consider unlikely given the timespan and quality of our data. Third, the tidal synchronization in binary $A$ has not completely finished due to either young age or continuous decrease of $e_A$ and at least one of the stars is rotating slightly faster than the pseudo-synchronous rate. Fourth, it is possible that three consecutive sectors of TESS data are insufficient to measure $f_\text{psrot}$ with this precision and the offset in frequencies is purely due to low signal to noise ratio. We consider explanations one and four, or their combination, most likely, but in Section~\ref{sec:past} we will briefly explore the implications of stars in binary $A$ rotating faster  than the pseudosynchronous rate.

Finally, we identified a mode with a period of about $0.45$\,days, which is not easily matched to harmonics of any obvious frequency in the system. This mode could be caused by pulsations of one the stars. This does not affect any of our subsequent analysis so we do not discuss this finding anymore.

\subsection{Apsidal motion}
\label{sec:apsidal}

\renewcommand{\arraystretch}{1.5}
\begin{table*}
\caption{Apsidal motion contributions for binary $A$. }              
\label{tab:aps}      
\centering                                      
\begin{tabular}{l c c c}          
\hline\hline                        
Contribution & star 1 & star 2 & total \\    
\hline                                   
$\displaystyle \dot{\omega}_\text{GR} = \frac{2\pi}{P_A} \frac{3GM_A}{c^2 a_A (1-e_A^2)}$ &  & & $1.47\times 10^{-5}$ \\[5pt]
$\displaystyle \dot{\omega}_{\text{tide},l} = \frac{30\pi k_{2,l}}{P_A} \left(\frac{M_{A,3-l}}{M_{A,l}} \right) \frac{8 + 12e_A^2 + e_A^4}{8\left(1-e_A^2\right)^5} \left(\frac{r_l}{a}\right)^5$ & $2.00 \times 10^{-4}$ & $2.14\times 10^{-4}$ & $4.14 \times 10^{-4}$ \\
$\displaystyle \dot{\omega}_{\text{rot},l} = \frac{2\pi k_{2,l}}{P_A} \left(\frac{P_A}{P_\text{rot}}\right)^2 \frac{1 + M_{A,3-l}/M_{A,l}}{\left(1-e_A^2\right)^2}\left(\frac{r_l}{a}\right)^5$ & $2.58\times 10^{-5}$ & $3.81\times 10^{-5}$ & $6.40\times 10^{-5}$\\
$\displaystyle \dot\omega_{AB} = \frac{15\pi}{4} \frac{M_B}{M_A+M_B} \frac{P_A}{P_{AB}^2} \frac{(1-e_{AB}^2)^{-3/2}}{(1-e_A^2)^{1/2}} \mathcal{F}(\cos i_{A,AB})$ & & &  $-1.70\times 10^{-6}$\\
\hline
Total & & &  $4.90 \times 10^{-4}$\\
\hline
Observed & & &  $5.13 \times 10^{-4}$\\
\hline
\end{tabular}
\tablefoot{All values are given in rad/day and we assumed $\phi_l = 1$ for both components. Equations come from \citet{liang22} and \citet{borkovits15}. The function $\mathcal{F}$ is defined in Eq.~(C8) of \citet{borkovits15} and depends on the mutual inclination of orbit $A$ to the outer orbit. Its value is about $-0.20$ for mutually perpendicular orbits ($\cos i_\text{A,AB} = 0$) compatible with constraints from Sect.~\ref{sec:mutual}.}
\end{table*}

Binary $A$ shows a clear apsidal motion with a period of $2\pi/\dot{\omega} \approx 33.4^{+0.2}_{-0.2}$\,years (Fig.~\ref{fig:oc_orig}, Tab.~\ref{tab:fit}). To understand the source of the apsidal motion, we evaluate the individual contributions following, for example, \citet{philippov13} and \citet{liang22}. The apsidal motion is composed of general relativity ($\dot\omega_\text{GR}$), tidal ($\dot\omega_\text{tide}$), rotational ($\dot\omega_\text{rot}$), and outer orbit ($\dot\omega_{AB}$) components:
\begin{equation}
    \dot{\omega} = \dot{\omega}_\text{GR} + \sum_l \dot{\omega}_{\text{tide},l}+ \sum_l \dot{\omega}_{\text{rot},l}\phi_l + \dot\omega_{AB},
\end{equation}
where the summation is over both stars, $l=\{1,2\}$, and $\phi_l$ depends on the spin orientation relative to the orbit. When spins and orbit are perfectly aligned, $\phi_l = 1$.

In Table~\ref{tab:aps}, we present the expressions for individual contributions to apsidal motion and their numerical values for individual components of binary $A$. To estimate the values of apsidal motion constant, we use the tables of \citet{claret04} and investigate the range of $k_2$ for models with initial masses similar to our fiducial values in Table~\ref{tab:my_label} and with age between $300$\,Myr and 1\,Gyr. Specifically, we find that $-2.09 \le \log k_{2,1} \le -2.05$ and  $-2.54 \le \log k_{2,2} \le -2.41$. We choose values approximately in the middle of the intervals, specifically $\log k_{2,1} = -2.07$ and $\log k_{2,2} = -2.47$. We see that apsidal motion is dominated by tidal and rotational contributions, while general relativity and perturbations from the other binary are very small. The observed rate is very close to the theoretical value with the assumption that the spins are aligned with the orbit. We could get a perfect agreement if we slightly moved the values $k_2$ within the ranges quoted here, for example, $\log k_{2,1} = -2.06$ and $\log k_{2,2} = -2.45$ give total apsidal motion rate of $5.08\times 10^{-4}$\,rad/day. It is also worth pointing out that if masses of stars in binary $A$ were higher (Sect.~\ref{sec:physical}), then the theoretical apsidal precession rate could be significantly lower. For example,  if the mass of star $1$ in binary $A$ was $1.41\,\msun$, tables of \citet{claret04} predict $\log k_{2,1} \approx -2.32$, which would lower the total apsidal precession rate to about $4.0 \times 10^{-4}$\,rad/day. This could be viewed as a support for lower mass solution in CzeV343.

It is important to point out that the observed apsidal rate can be substantially different from the dynamical rate observed in the frame of the binary, which can be precessing either due to misaligned spins or an external perturbing object. \citet{philippov13} analyzed this scenario for DI~Her, where the spins and the orbit are misaligned. \citet{borkovits07} focused on triple stellar systems and found that the measured $\dot{\omega}$ can be smaller than the theoretical value if only a fraction of the apsidal cycle is covered with observations and if the outer orbit is oriented almost in the plane of the sky. As we will show in Section~\ref{sec:mutual}, this is exactly the situation in CzeV343. We find relatively good match between the observed and theoretically-predicted values of $\dot{\omega}_A$, but these complications should be kept in mind.

Finally, eccentricity of binary $B$ in our photometric model, $e_B = 0.0017_{-0.0013}^{+0.0020}$, is consistent with a circular orbit, which makes any estimates of apsidal motion impossible at this point. 

\subsection{Mutual orbit}
\label{sec:mutual}

As part of our photometric model we included LTTE due to the mutual orbit of binaries $A$ and $B$ and thus prove that they are bound. We find that the measured LTTE amplitude is only $D_A \approx 1.75\times 10^{-3}$\,days. From Equation~(\ref{eq:ltte}) and using our values for $e_{AB}$ and $\omega_{AB}$, we would expect amplitude $8.60_{-1.3}^{+0.8} \times 10^{-3} \sin i_{AB}$\,days. This implies that $\sin i_{AB} \approx  0.205_{-0.015}^{+0.026}$ and that the inclination of the mutual orbit is only about $ 11.8_{- 0.9}^{+1.5}$\,degrees. These estimates include only the uncertainties on $D_A$, $P_{AB}$, $e_{AB}$, and $\omega_{AB}$ and not on stellar masses. However, the observed LTTE amplitude is so small that the mutual orbit must be nearly face-on for any realistic stellar masses.

Based on this inference, we can estimate the range of inclinations between binaries $A$ and $B$ and their mutual orbit, $i_{A,AB}$ and $i_{B,AB}$. From LTTE, we know that the angular momentum vector of the mutual orbit is nearly perpendicular to the plane of the sky and that the angular momentum vectors of the two inner binaries are in the plane of the sky.  Therefore, $i_{A,AB}$ and $i_{B,AB}$ must be large. The angle $i_{A,AB}$ must satisfy
\begin{equation}
    \cos( i_A + i_{AB}) \le \cos i_{A,AB} \le \cos (i_A - i_{AB}),
\end{equation}
where we need to keep in mind the $180^\circ$ ambiguity of the angles. A similar relation holds for $i_{B,AB}$. We find that $76^\circ \lesssim i_{A,AB} \lesssim 104^\circ$ and either $63^\circ \lesssim i_{B,AB} \lesssim 86^\circ$ or $94^\circ \lesssim i_{B,AB} \lesssim 117^\circ$. We verified these constraints by Monte Carlo simulation of unit vector distributed on a sphere. Our results imply that even three-dimensional orbits are highly inclined with respect to each other. We discuss evolutionary implications for this misalignment in Section~\ref{sec:past}.

Our finding of misalignment in CzeV343 make this object potentially unique among other DEBs. However, it is important to point out that a randomly oriented orbit only has a small probability of lying in the plane of the sky. It is more likely that the orbit will be significantly inclined. Since with LTTE we cannot constrain mutual orientation of DEB orbits when viewed edge-on, it cannot be excluded that many DEBs actually have significantly misaligned orbits. This problem can be addressed with larger samples of fully characterized DEBs.

\subsection{Nodal precession}
\label{sec:nodal}

The large mutual inclination between inner and outer orbits will lead to nodal precession of inner orbits. This has three consequences. First, the apsidal motion will include additional term due to nodal precession. \citet[Appendix~C]{borkovits15} gives expression for contribution to the apsidal rate from nodal precession. Its magnitude is similar to $\dot{\omega}_{AB}$ given in Table~\ref{tab:aps} and therefore its effect will be similarly small. 

Second, observed inclinations of the inner binaries will change over time, which can be observed as changing eclipse depths. The typical timescale for inclination change is $P_{AB}^2/P_A \approx 4800$\,years. This is much longer than the span of observations we might hope to achieve. However, the eclipse depth can be very sensitive to inclination changes, for example, for grazing eclipses. Our fitting code does not straightforwardly allow for fitting of changing inclination. However, looking at the fit of individual nights over 11\,years of our data (Figs.~\ref{fig:ground_lc1}--\ref{fig:ground_lc2}), we do not see any trends in minima depths that are not captured by the photometric model. Perhaps these changes will be detected in the future.

Third, nodal precession can introduce non-trivial interaction between the spins and orbits of the binaries. To investigate the short-term spin-orbit dynamics, we implemented the equations of \citet{eggleton01} taking into account GR and quadrupolar distortion precession, perturbations by a distant body, and neglecting all tidal interactions. We set up the calculation to resemble binary $A$, which is perturbed by a distant companion with mass equal to the total mass of binary $B$. We see fast apsidal motion and significantly slower nodal precession, as expected. We do not see any change of angle between the spin and orbital angular momentum irrespective of whether the spin started aligned or misaligned relative to the orbit. Since we observe synchronous rotation in $A$, the tides have also quite likely aligned the spins. The situation was different in the past when the orbit was wider and apsidal precession rate significantly smaller. We model this scenario by simply setting $k_2$ to $10^{-4}$ of their original value. We observe dramatic changes of spin-orbit angle on the timescale of nodal precession. This effect might be observable in 2+2 quadruples with inner orbits wider than in CzeV343.

\subsection{Period resonance}
\label{sec:resonance}

\begin{figure*}
    \centering
    \includegraphics[width=0.9\textwidth]{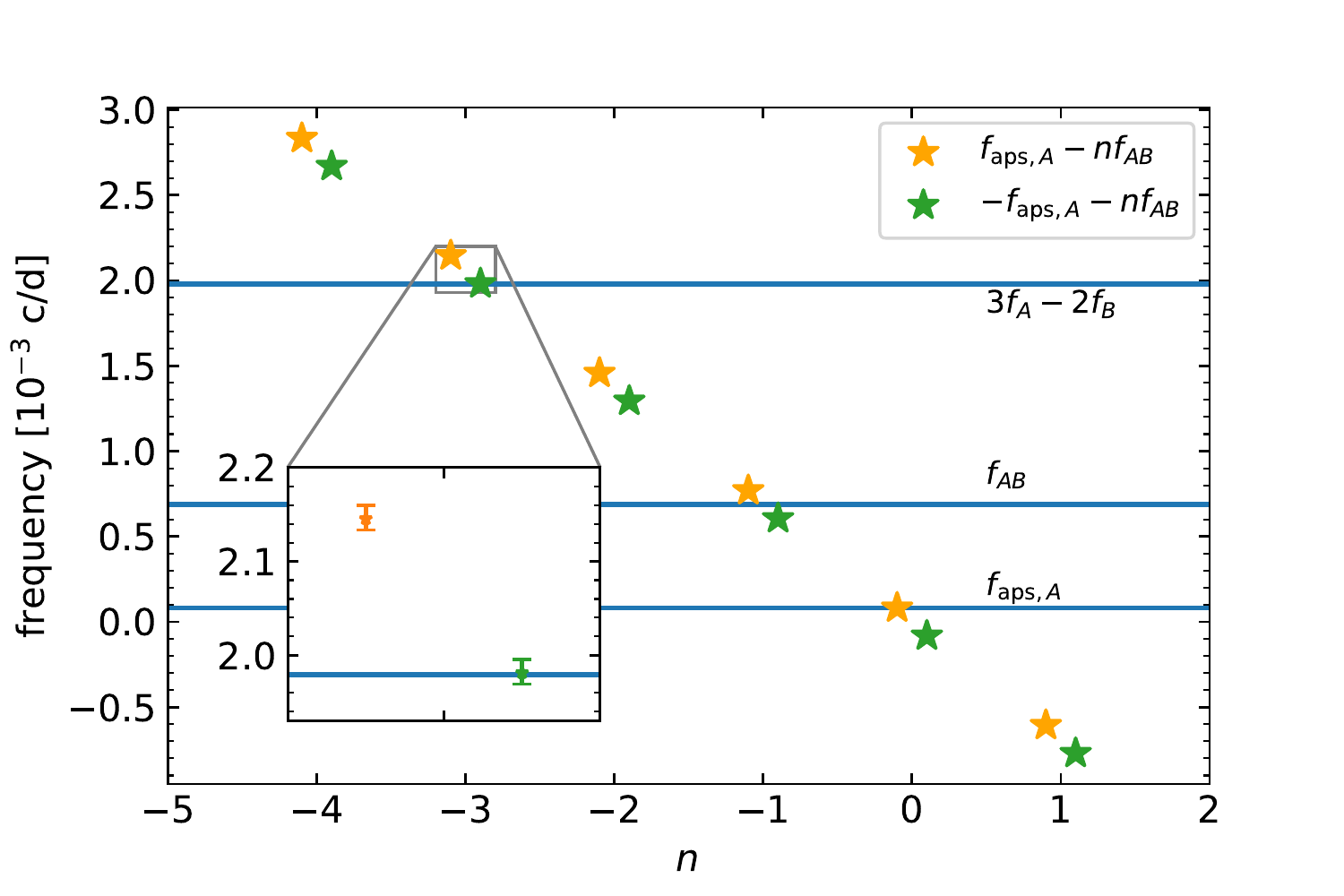}
    \caption{Period resonance in CzeV343. Blue horizontal lines show individual terms from Eq.~(\ref{eq:resonance}) as determined by our photometric model. Orange and green stars show combinations of $f_{\text{aps},A}$ and $f_{AB}$ for different values of $n$. The inset plot shows zoom-in for $n=-3$, where we also show $95.4\%$ confidence intervals from our model.}
    \label{fig:resonance}
\end{figure*}

The most striking feature of CzeV343 is the close period resonance. Our updated photometric model gives
\begin{equation}
    3f_A - 2f_B \approx (1.9794 \pm 0.0003) \times 10^{-3}\ \text{cycles/day},
\end{equation}
which looks significantly far away from the resonance. However, \citet{tremaine20} showed that the 3:2 resonant condition includes also apsidal motion and mutual orbit as
\begin{equation}
3f_A - 2f_B = f_{\text{aps},A} - n f_{AB},
\label{eq:resonance}
\end{equation}
where $f_{\text{aps},A} = \dot{\omega}_A/(2\pi)$, and $n$ is a small integer. It is important to point out that this resonant condition was derived for exactly aligned inner and outer orbits, which is dramatically different from the highly misaligned orbits in CzeV343. It is also possible that the resonant mechanism is different for misaligned orbits. Furthermore, the apsidal rate $\dot{\omega}$ seen from the outer orbit is potentially different from the true rate in the frame of the inner binary, as we discussed in Section~\ref{sec:apsidal} \citep{borkovits07,philippov13}. Despite these complications, it is still instructive to evaluate Equation~(\ref{eq:resonance}) with the observed values of the relevant quantities.

But before we do that, we need to make clear whether our observed values and Equation~(\ref{eq:resonance}) are in the same frame of reference. The orbital parameters can be formulated either in the sidereal frame, where angles are measured relative to the distant observer, or in the anomalistic frame, where angles are measured relative to the orbital pericenter, which can be precessing in time. As a result, the resonant conditions in these two frames should be related by the apsidal precession frequency. Orbital periods in Table~\ref{tab:fit} are in the sidereal frame, because we are measuring the mean interval between eclipses. The theoretical condition is also in the sidereal frame (Tremaine, 2022, private communication). We can therefore directly compare our observations with the theoretical condition\footnote{ A more thorough explanation of the structure of resonant relations in the context of precessing orbits in the Solar system is given by \citet[Chapter 8.2]{murray}. Discussion of sidereal and anomalistic frames of reference in the context of apsidal motion in binary stars is given, for example, by \citet{gimenez83}.}.

In Figure~\ref{fig:resonance}, we show with orange stars the individual terms of Equation~(\ref{eq:resonance}) and their combinations for different $n$. We see that the best match for the period resonance is achieved for $n=-3$, but there is still disagreement of about $10^{-4}$\,cycles/day, which is much larger than our formal confidence interval. We already discussed that MCMC exploration gives surprisingly small uncertainties on the properties of the mutual orbit. If we increased the uncertainty of $P_{AB}$ to $10$ or $20\%$, the resonance would agree with the uncertainties.

However, we found out that if we change the sign in front of $f_{\text{aps},A}$ in Equation~(\ref{eq:resonance}) as 
\begin{equation}
3f_A - 2f_B = -f_{\text{aps},A} - n f_{AB},
\label{eq:resonance_alt}
\end{equation}
the match becomes essentially exact as we show with green stars in Figure~\ref{fig:resonance}. The difference between left- and right-hand sides of Equation~(\ref{eq:resonance_alt}) becomes only $-0.2^{+1.5}_{-1.2} \times 10^{-5}$\,cycles/day. It is not clear how to interpret this finding, but it might represent a different resonance condition for misaligned binaries or some kind of its projection due to geometrical effects mentioned in Section~\ref{sec:apsidal}.

\section{Discussions}
\label{sec:disc}

In this section, we discuss evolution of CzeV343 focusing on how it achieved the current resonant state (Sect.~\ref{sec:past}) and how it will evolve in the future (Sect.~\ref{sec:evol}).

\subsection{Past evolution}
\label{sec:past}

\citet{tremaine20} presented several conditions that need to be satisfied for capturing 2+2 quadruples in the 3:2 resonance. This theory assumes that all orbits are coplanar, which is violated in CzeV343. Although the actual resonance mechanism for misaligned orbits might be different from coplanar orbits, it is still interesting to discuss the coplanar resonant conditions in the context of CzeV343. First, the theoretical range of mean motions due to resonance splitting is
\begin{equation}
\left|1- \frac{2P_A}{3P_B} \right| = 0.47|n|\epsilon^{3/2} \approx 8.6 \times 10^{-4},
\end{equation}
where $\epsilon = a_A/a_{AB} \approx 7.2 \times 10^{-3}$ is the small parameter in the system Hamiltonian, and we set $|n|=3$ as derived in Section~\ref{sec:resonance}. From our photometric model, we find $|1-2P_A/3P_B| \approx 7.9790 \pm 0.0012 \times 10^{-4}$ and CzeV343 therefore fits within the resonance width, as \citet{tremaine20} also concluded.

Second, capture to coplanar resonance requires that orbital frequencies evolve as
\begin{equation}
    \frac{\dot{f_A}}{f_A} > \frac{\dot{f_B}}{f_B}.
    \label{eq:fevol}
\end{equation}
Since all stars in CzeV343 likely have radiative envelopes, magnetic braking is not efficient and the binary separations should evolve by gravitational radiation as
\begin{equation}
    \frac{\dot{f}}{f} = \frac{96G^3M_1M_2(M_1+M_2)}{5c^5 a^4} = \begin{cases}
    2.3\times 10^{-19}\,\text{s}^{-1} & \text{binary}\ A, \\
    5.3\times 10^{-19}\,\text{s}^{-1} & \text{binary}\ B ,
    \end{cases}
\end{equation}
which implies that Equation~(\ref{eq:fevol}) is not satisfied. Star 1 in binary $A$ and both stars in binary $B$ are actually very close to the dividing line between convective and radiative envelopes and magnetic braking could have been operating in CzeV343. However, since binary $B$ is closer and has lower mass, its magnetic braking would be stronger than in $A$ and its evolutionary timescale due to magnetic braking would be shorter than for binary $A$, again not satisfying the requirement of Equation~(\ref{eq:fevol}).

Third, capture to coplanar resonance requires that binary $A$ has a nearly circular orbit
\begin{equation}
    e_A \lesssim \epsilon^{5/3} \approx 2.7 \times 10^{-4},
    \label{eq:ecc}
\end{equation}
which is not satisfied by a large margin. \citet{tremaine20} proposed that eccentricity in DEBs can be excited after the resonant capture. For CzeV343, the obvious candidate are von Zeipel--Lidov--Kozai (vZLK) cycles, because we showed that the orbits are currently highly inclined. \citet{antognini15} and \citet{naoz16} give the timescale for vZLK cycles as
\begin{equation}
    t_\text{vZKL} = \frac{16}{30\pi} \frac{M_A+M_B}{M_B} \frac{P_{AB}^2}{P_A} (1-e_{AB}^2)^{3/2} \approx 650\,\text{years}.
    \label{eq:kozai}
\end{equation}
This timescale is about an order of magnitude longer than the apsidal precession period in binary $A$, which is primarily set by tidal and rotation contributions (Sect.~\ref{sec:apsidal}). This makes vZLK inoperable in the current configuration of CzeV343 \citep[e.g.][]{fabrycky07}. vZLK cycles could have operated earlier in the evolution CzeV343 when the orbits were wider and apsidal precession was less important.

What is the origin of eccentricity in binary $A$? \citet{justesen21} calculated theoretical circularization orbital periods for binaries of various effective temperature based on models of \citet{claret04}. For binaries with effective temperatures similar to CzeV343, the critical circularization orbital period is $1$--$2$\,days. Since the circularization is a strong function of stellar radii relative to orbital size and $(r_1+r_2)_B > (r_1+r_2)_A$, it is highly likely that binary $B$ has already circularized while binary $A$ is still undergoing its circularization process. This implies that the circularization timescale $t_\text{circ}$ of binary $A$ is about the age of system, which is a few hundred Myr. However, \citet{justesen21} identified many hot binaries with longer periods and circular orbits, which suggests that tides can be, at least in some situations, significantly more efficient than what dynamical theory predicts. It is thus not impossible that $t_\text{circ}$ in binary $A$ is significantly shorter than the age of the system.

\begin{figure*}
    \centering
    \includegraphics[width=\textwidth]{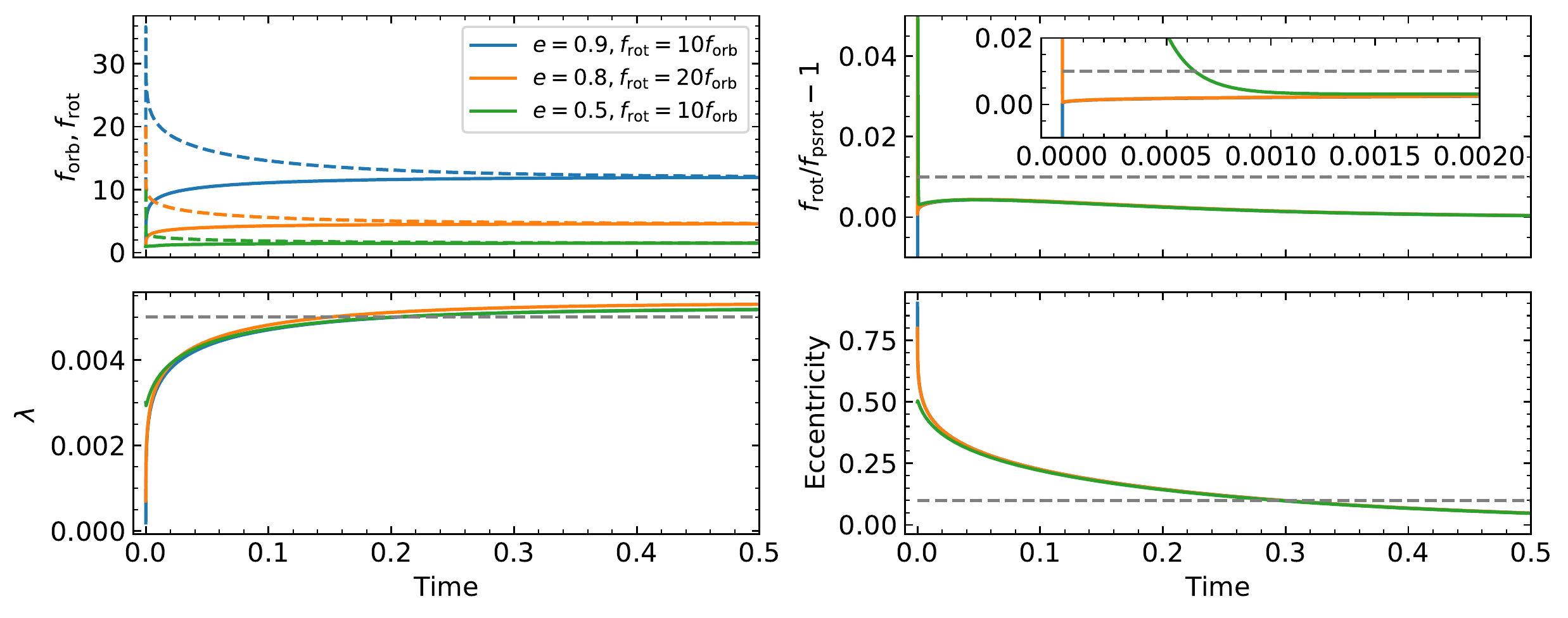}
    \caption{Several examples of possible tidal evolution of binary $A$ in CzeV343. In the top left panel, we show orbital (solid lines) and stellar rotation (dashed lines) frequencies for different initial parameters indicated in the legend. In the top right panel, we show the ratio between rotation and pseudo-synchronous frequencies including a zoom-in on the early part of the evolution. The vertical dashed line shows the $1\%$ excess possibly detected in CzeV343. In bottom left panel, we show the evolution of $\lambda$ that converges near the current estimate of $0.005$ marked with horizontal dashed line. In bottom right panel, we display the evolution of eccentricity with the value $0.1$ marked with a horizontal dashed line. Time is measured in the units of the tidal friction timescale, which is approximately $9t_\text{circ}$ \citep{eggleton06}.}
    \label{fig:tidal}
\end{figure*}

In Section~\ref{sec:rotation}, we detected a signal very close to the pseudo-synchronous rotation frequency of binary $A$. It is interesting to discuss what does this imply for the origin of the eccentricity. Synchronization timescale $t_\text{sync}$ is shorter than the circularization timescale by the ratio of spin to orbital angular momentum $\lambda$,  $t_\text{sync} = \lambda t_\text{circ}$ \citep{eggleton06}. For star $2$ in binary $A$, $\lambda$ is
\begin{equation}
    \lambda = r_\text{gyr}^2  r_{2,A}^2 \frac{M_A}{M_{1,A}}\frac{f_\text{rot}}{f_A} \approx 5\times 10^{-3},
    \label{eq:lambda}
\end{equation}
where $r_\text{gyr} \approx 0.2$ is the relative gyration radius based on models of \citet{claret04}, and $r_2$ is the relative radius from photometry (Tab.~\ref{tab:fit}). Pseudo-synchronous rotation in binary $A$ implies that the eccentricity was excited more than about $t_\text{sync}$ ago, which is about few Myr if $t_\text{circ}$ is approximately the age of the system. This is much longer than $t_\text{vZLK}$, but consistent with the fact vZLK are currently not operating in CzeV343.

The frequency in photometric residuals appears to be about $1\%$ faster than the pseudo-synchronous rotation frequency. There remains doubt whether this excess is real, but taken at face value, does this excess imply that synchronization in binary $A$ is still ongoing? To study this question, we have implemented a simple equilibrium tidal model for the evolution of orbital and stellar rotation frequencies and the eccentricity from \citet[Chap. 4.2]{eggleton06}. We manually select initial values of $\lambda$ so that the asymptotic values are similar to what we estimated for binary $A$ assuming that the absolute stellar radius is constant and $\lambda \propto f^{4/3}$. In Figure~\ref{fig:tidal}, we show several examples of the evolution. In all cases, the rotation frequency excess $f_\text{rot}/f_\text{psrot} -1$ converges to a small positive value, which is noticeably smaller than $1\%$. The frequency excess exceeds $1\%$ only during the fast initial synchronization period, but we would have to be extremely lucky to catch CzeV343 shortly after its $e_A$ was excited. Substantial eccentricity seems to be common in many DEBs, which argues against lucky eccentricity excitation.

To summarize our discussion, although CzeV343 is located nearly exactly on the 3:2 period resonance, many of its properties are not consistent with the theoretical requirements for resonant capture for coplanar orbits. This is not too surprising, because we showed that orbits in CzeV343 are highly misaligned. We think that the most plausible scenario is that at or shortly after its birth, CzeV343 reached a state, where binary $A$ had substantial eccentricity but binary $B$ was already close to its current configuration. Eccentricity in binary $A$ could have been excited by vZLK oscillations at times when $A$ was wider and the intrinsic apsidal precession was unimportant. The subsequent tidal evolution synchronized the rotation and nearly circularized the orbit while decreasing the semi-major axis. This scenario satisfies the condition in Equation~(\ref{eq:fevol}), but violates the requirement of small eccentricity (Eq.~[\ref{eq:ecc}]) and that the  eccentricity damping timescale must be much shorter than the semi-major axis evolution timescale \citep{tremaine20}. Clearly, resonance capture theory that takes into account misaligned orbits and vZLK cycles would be very useful for explaining CzeV343.

\subsection{Future evolution}
\label{sec:evol}

We are interested to learn about the future evolution of CzeV343. Leaving aside the unknown consequences of period resonance, we could use the code MSE, which combines secular and dynamical evolution of multiple stellar systems with stellar evolution \citep{hamers_mse}. Similar study was performed, for example, by \citet{merle22}. However, the ratio of orbital periods between the inner and outer orbits in CzeV343 is so large that individual model runs in MSE would take many tens of hours. We would need many such runs, because we cannot fully constrain the mutual geometry of the orbits. Instead, we focus on the evolution of individual binaries and discuss the influence of their mutual orbit only qualitatively. We use the binary population code COMPAS \citep{compas} to synthesize a number of possible realizations of binaries of binaries $A$ and $B$.  We use the default parameters for binary evolution.

Given the uncertainties in stellar parameters of CzeV343 and in the theoretical models, we explore a range of initial parameters. For binary $A$, our grid covers $1.6 \le M_{A,2} \le 1.9\,\msun$, $1.0 \le P_A \le 1.4$\,days, $0.5 \le q_A \le 0.9$, and $0 \le e_A \le 0.15$. We find that only about $20\%$ of such binaries survive, mostly as binary He and C/O white dwarfs. The remaining binaries merge either on the main sequence or when one star is a He white dwarf and the other is on the helium Hertzsprung gap. We find that $\gtrsim 85\%$ of simulated binaries experience significant orbital expansion, when the initially more massive star evolves to very low masses by mass transfer to its companion. The maximum semi-major axis achieved in the evolution ranges from $50$ to $220\,\rsun$. In such situations, the entire quadruple system could become dynamically unstable. This would be very similar to the triple evolution dynamical instability (TEDI) triggered by orbital expansion driven by mass transfer \citep{perets12,hamers22}. We do not have stability criteria for quadruples, but we can approximate the system as a triple by replacing the other binary with a point mass with the same total mass. In such case, we can use the stability criterion of \citet{mardling01}
\begin{equation}
    \frac{a_{AB}}{a_A} > 2.8 \left(1 + \frac{M_B}{M_A}\right)^{2/5} \frac{(1+e_{AB})^{2/5}}{(1-e_{AB})^{6/5}},
\end{equation}
which gives $a_{AB}/a_A \gtrsim 19$ for the parameters of the mutual orbit and binary masses from Table~\ref{tab:my_label}. This implies that only orbits with $a_A \lesssim 50\,\rsun$ are dynamically stable and that essentially all realizations of binary $A$ that survive main sequence as a binary will experience this dynamical instability.

For binary $B$, we run models with $1.2 \le M_{B,1} \le 1.5\,\msun$, $0.75 \le q_B \le 1.0$, $0.6 \le P_B \le 1.0$\,days, and $e_B=0$. We find that apart from several runs that produce double He white dwarfs, vast majority of the simulations lead to mergers either of main sequence stars or helium stars. Similarly to binary $A$, about $60\%$ of realizations experience orbital expansion to between $80$ and $150\,\rsun$. Applying the stability condition of \citet{mardling01}, we find that the binary $B$ is stable also only for $a_B \lesssim 50\,\rsun$ and again a large fraction of realizations will end up dynamically unstable.

Finally, we explore what happens when either or both of the binaries merge already on the main sequence. To estimate the range of possible outcomes, we neglect any changes in the structure and mass of the merger remnants\footnote{Analysis of recent mergers from low-mass stars suggests that only about $10\%$ of the binary mass is lost during the interaction \citep[e.g.][]{nandez14,pejcha17,macleod17}.} and approximate the system as a binary with primary mass $2.7 \le M_A \le 3.3\,\msun$, mass ratio $0.7 \le q_{AB} \le 1.0$, orbital period $1300 \le P_{AB} \le 1500$\,days, and eccentricity $0.6 \le e_{AB} \le 0.8$. We find that about $85\%$ of realizations evolve to C/O binary white dwarfs. About $10\%$ end as Type Ia supernova explosion and the rest are either main sequence mergers or double white dwarf binaries composed of He and C/O white dwarfs. We note that if the masses in CzeV343 are actually higher than what is assumed here (Tab.~\ref{tab:my_label}), the outcomes associated with more massive stars such as Type Ia supernovae would become more likely. Since binary $A$ is more massive than binary $B$, it would merge first and the remnant can evolve to red giant while binary $B$ is still near the main sequence. In this case, some of these evolutionary pathways would actually include mass transfer from the merged product of binary $A$ to binary $B$. This mass transfer can either remain stable or it can culminate in triple common envelope evolution \citep{glanz21,hamers22_triple,merle22}. 

In Sect.~\ref{sec:physical}, we could not unambigously establish the temperatures and masses of the stars. It remains possible that the individual components are more massive than our fiducial assumption. We expect that fractional probabilities of the individual evolutionary pathways would change. In particular, we expect to have a higher frequency of binaries ending as Type Ia supernovae. However, the total mass of the quadruple is still smaller than $8\,\msun$ and it is unlikely that the evolution would end with a formation of a neutron star or a black hole.


\section{Conclusions}
\label{sec:conc}

We have analyzed long-term ground-based optical photometry together with TESS 2-minute photometry and two optical spectra of CzeV343. Our results can be summarized as follows.
\begin{itemize}
    \item We construct a photometric model for the two eclipsing binaries on a mutual orbit manifested by LTTE (Sect.~\ref{sec:global}). We also include possibility of an additional light in the system and we take into account instrumental signals simultaneously with the physical model. We explore the parameter space with MCMC and we constrain all of the model parameters (Table~\ref{tab:fit}, Fig.~\ref{fig:lc_comp}, Appendices~\ref{app:ground}, \ref{app:tess}, \ref{app:posteriors}). 
    \item By combining relative radii from photometry, relative amplitudes of LTTE, spectral type based on one of our optical spectra, and theoretical single-star isochrones, we constrain physical properties of all four stars in CzeV343 (Sect.~\ref{sec:physical}). We find that binary $A$ is composed of stars of approximately $1.76$ and $1.27\,\msun$ and binary $B$ has stars of mass of approximately $1.36$ and $1.20\,\msun$ (Tab.~\ref{tab:my_label}). The age of the system is a few hundred Myr (Fig.~\ref{fig:iso}). The uncertainties on our results are considerable and complex. Radial velocity study of CzeV343 would be beneficial, but would be difficult due to wide absorption lines in spectrum.
    \item We perform period analysis on the residuals of our photometric model and identify a signal consistent with pseudo-synchronous rotation in binary $A$ (Sect.~\ref{sec:rotation}). The rotation signal is possibly about $1\%$ faster than the pseudo-synchronous rate, but the duration of the TESS 2-min data is too short to determine this more exactly (Fig.~\ref{fig:period}). Analysis of stellar rotation in DEBs is an unexplored territory, which could give new clues to their formation.
    \item We see a clear apsidal motion in binary $A$ with a period of $33.4 \pm 0.2$\,years (Sect.~\ref{sec:apsidal}). We calculate contributions to the apsidal precession rate from general relativity, tidal deformation, and rotation, and find that the observed rate can be fully explained assuming the stellar spins are aligned with the orbit (Tab.~\ref{tab:aps}). Interpretation of observed apsidal motion is potentially complicated by orbital precession induced by binary $B$. Eccentricity of binary $B$ is consistent with zero.
    \item Our photometric model gives for the orbital period of the mutual orbit of the two binaries as $P_{AB} \approx 1454^{+10.2}_{-8.3}$\,days and its eccentricity as $e_{AB} \approx 0.70^{+0.09}_{-0.07}$, which proves that the two binaries are gravitationally bound (Sect.~\ref{sec:mutual}). The amplitudes of the LTTE of both binaries are significantly smaller than what would be expected based on their masses, which suggests that the outer orbit has a low observed inclination of about $11^\circ$. By randomly sampling possible orientations of the individual orbits, we find that the inner and outer orbits are significantly misaligned and potentially even retrograde. The constraints on the relative inclinations of the $A$ and $B$ orbits and the outer orbit are $76^\circ \lesssim i_{A,AB} \lesssim 104^\circ$ and either $63^\circ \lesssim i_{B,AB} \lesssim 86^\circ$ or $94^\circ \lesssim i_{B,AB} \lesssim 117^\circ$.
    \item We quantify the 3:2 period resonance of the two inner orbits (Sect.~\ref{sec:resonance}). We find that if we significantly increase the uncertainty in $P_{AB}$ to about $10\%$, we can match the theoretical combination of orbital and apsidal frequencies derived by  \citet{tremaine20} (Eq.~[\ref{eq:resonance}]). However, if we change the sign of the apsidal motion (Eq.~[\ref{eq:resonance_alt}]), the resonance becomes nearly exact with our frequency uncertainty of $10^{-5}$ cycles/day (Fig.~\ref{fig:resonance}). 
    \item We find that the properties of CzeV343 are not compatible with many requirements of the resonance capture theory developed by \citet{tremaine20} for coplanar orbits (Sect.~\ref{sec:past}). This is not too surprising given the high misalignment of inner and outer orbits. The current eccentricity of binary $A$, $e_A \approx 0.11$, and synchronous rotation of at least one of its stars disfavor recent eccentricity excitation. Instead, we propose that binary $A$ is tidally circularizing from a past high-eccentricity state, which could have been caused by von Zeipel--Lidov--Kozai oscillations. Theory of resonant capture should be modified to take into account misaligned and eccentric orbits.
    \item We explore future evolutionary pathways of binaries $A$ and $B$ using binary population synthesis code COMPAS (Sect.~\ref{sec:evol}). We find that either binary typically merges on the main sequence or evolves to a double white dwarf binary. During mass transfer episodes, the mass ratio can become so extreme that the associated orbital expansion will trigger dynamical instability similarly to what was identified by \citet{perets12} and \citet{hamers22} for triples. However, detailed properties of this instability remain unknown in the case of 2+2 quadruples. If both binaries merge on the main sequence, the total mass of the system is sufficiently high and the mutual orbit is sufficiently compact to open the possibility of evolution leading to a Type Ia supernova. 
    \item In Appendix~\ref{app:sips}, we describe algorithms and functionality of computer programs SIPS and SILICUPS, which can help to efficiently organize and analyze extended time series photometry of variable stars and similar objects.
\end{itemize}

\begin{acknowledgements}
We thank the referee for insightful comments that improved the manuscript. We thank Scott Tremaine, Todd Thompson, Adrian Hamers, and Max Moe for discussions and comments that improved the paper.  PC thanks V\'{a}clav P\v{r}ib\'{i}k for initial implementation of the algorithm determining eclipsing binary periods from minima timings and Petr Caga\v{s} for implementation of the 2D polynomial regression used to model telescope field curvature.
 
 The work of OP, PC, CL, and MP has been supported by INTER-EXCELLENCE grant LTAUSA18093 from the Ministry of Education, Youth, and Sports. The work of OP has been additionally supported by Horizon 2020 ERC Starting Grant ‘Cat-In-hAT’ (grant agreement no. 803158). This research was supported in part by the National Science Foundation under Grant No. NSF PHY-1748958.
 
 This paper includes data collected by the TESS mission,
which are publicly available from the Mikulski Archive for
Space Telescopes (MAST). Funding for the TESS mission is
provided by NASA’s Science Mission directorate. Most of the calculations and visualizations in this work were performed with \textsc{matplotlib} \citep{hunter07}, \textsc{scipy} \citep{scipy}, and \textsc{numpy} \citep{harris20}. 
\end{acknowledgements}

\bibliographystyle{aa} 
\bibliography{mybib} 

\begin{appendix} 
\section{Description of SIPS \& SILICUPS}
\label{app:sips}

Astronomical instrumentation commercially available to citizen scientists is quickly advancing. The trend is especially prominent with imaging detectors, where improvements in size and sensitivity enable new projects. Specifically,  taking measurements of a single star in an entire image is being replaced by projects that analyze light curves of all stars in the image with the hope of discovering new variables or unexpected features in the already known ones. The mode of operation of many citizen skywatchers begins to resemble professional time-domain surveys. But there is an important difference: while professional surveys take one or at most few points every night, citizen scientist can easily take few hundred images of the same field every night. 

These advances bring new challenges related to data processing, where manual operations with  a single time series are simply not feasible anymore. For example, ground-based data of CzeV343 were taken with a 16 Mpx camera with a $1.5\times1.5$\,deg$^2$ field of view. About $34\,000$ stars were regularly detected on each image and about $200$ objects are regularly monitored with $O-C$ diagrams, phased light curves, and other tools. In denser Milky Way fields, the number of stars can exceed $100\,000$ with correspondingly higher number of interesting objects. Clearly, this data volume and complexity requires appropriate tools that can be efficiently used by citizen scientists with little to no scripting experience and time to develop their own pipelines. Here, we give a brief description of two software packages for Windows operating system\footnote{Both programs are freely available at \url{https://www.gxccd.com/cat?id=146&lang=409}}, which efficiently utilize multi-core architecture of current processors and are controlled by graphical user interface.

\subsection{SIPS (Simple Image Processing System)}

SIPS functions cover two main areas. First, SIPS controls the imaging device and other observatory equipment (telescope mounts, filter wheels, focusers, observatory domes, and others), and second, SIPS performs image calibration and astrometric and photometric reduction.
Photometry processing consists of several steps that are described in more detail below: detection of stars, image alignment and astrometry, aperture photometry, processing of the results, and identification of variable objects.

\subsubsection{Detecting stars}

An algorithm for detection of stars was developed from scratch specifically for SIPS. The performance of the algorithm is controlled by four user-defined parameters:

\begin{itemize}
    \item Aperture defining  a box size, where the mean and standard deviation of pixel values are calculated. This box is moved through the image.
    \item Threshold above which a pixel is tested if it is a part of a stellar image. Threshold is specified in the number of standard deviations above the mean.
    \item A minimal number of neighboring pixels that must also satisfy the above-defined threshold. Neighboring pixels are 4 pixels adjacent in the $x$ and $y$ axes directions plus the 4 corner pixels. Up to 8 pixels are tested to lie above the calculated threshold.
    \item Number of iterations to refine the stellar mask.
\end{itemize}

A pixel is considered to be part of a star image only if its value lies above the specified threshold and the minimal defined number of adjacent pixels also pass the same condition. However, a group of such adjacent pixels is considered as a star only if no pixel of the group touches the border of the current box. This eliminates multiple detections of the same group at different box positions. Such a group will be detected as a star when the box is positioned within the image to contain all pixels fulfilling the criteria defined above. Additionally, a single bit-depth mask of the same size as the examined image (bitmask) is created for every image, where the bits correspond to pixels belonging to the stellar image. When the aperture box moves through the image, pixels already set in the bitmask are skipped.

Centroids (averaged positions weighted by pixel value above the background) of the continuous groups of pixels are used as the first estimate of stellar positions. Such positions can be affected by differences in the aperture box mean and standard deviation values if another star or its part lies within the box and if such star is not yet detected (its pixels are not flagged in the bitmask) during the first pass through the image. To avoid this, the algorithm performs a user-defined number of passes through the image to iteratively refine star positions.

The existing bitmask is used to eliminate pixels of already detected nearby stars within the aperture box. Eliminating star pixels leads to a more precise calculation of the box mean and standard deviation and therefore centroids of newly found stars have better precision. Furthermore, the bitmask is updated after each iteration. An irregular group of neighboring pixels representing each star detected during the previous pass is replaced by circles with centers on newly determined centroids and with radii determined from aperture statistics.

The algorithm used to determine  radii of stars scans the profile from the centroid in the four directions (up, down, left, and right) and measures the distance from the centroid to the first pixel lying below the user-defined threshold.  Typically, values from 1 to 2$\sigma$ are used, but the optimal value depends on the stellar profile, which can be affected by many factors such as quality and type of optics, focusing precision, tracking precision, seeing, etc. Typically, two iterations of bitmask refinements are enough, but the number of iterations is also a user-defined parameter.

It might happen that automatically determined radii of nearby stars overlap. This is not desirable especially when automatic apertures are used to calculate the fluxes. SIPS detects and eliminates overlapping radii and calculates new radii as the fraction of the Euclidean distance of star centroids corresponding to the ratio of the fluxes of the two stars.

This method of detecting stars requires an aperture big enough to contain even the brightest stars within the image. An aperture too small leads to omissions of bright stars. Conversely, an aperture too large can miss faint stars very close to bright ones, because  the statistics of the aperture box are affected by the bright star so much that the pixels of a nearby faint star remain under the threshold. The solution to this problem is an optional usage of two apertures. User can choose a small aperture, which ensures reliable finding of even the faintest stars in the image, even if they are close to bright stars. Subsequently, a second much bigger aperture is  used to detect the brightest stars within the image. Usage of two apertures is, of course, slower and it is recommended mainly for dense stellar fields (typically within the Milky Way).

\subsubsection{Mutual image alignment}

Alignment calculation is based on searching for similar triangles among individual images. Triangles are created from the user-defined number $N$ of the brightest stars, so $\binom{N}{3}$ triangles are used. The similarity of triangles is determined by two numbers representing the ratios of the longest triangle side to the other two sides. The similarity limit is a user-defined parameter. When similar triangles are found, a preliminary 2D transformation $2\times3$ matrix, representing rotation, translation, and scaling, is created using the least-squares method. The validity of such transformation is then tested on the brightest stars used to create triangles: if at least $50\%$ of stars can be matched, the transformation is accepted. To further refine the transformation, the least-squares method is applied on all matched stars to determine the final transformation matrix. Transformation matrices relative to the selected reference frame are then calculated for all images in the set.

\subsubsection{Astrometry}

Absolute astrometric solution for all images is optional in SIPS, however, many advanced SIPS functions rely on astrometry. SIPS currently supports four star catalogs (UCAC4, UCAC5, USNO-A2.0, and USNO-B1.0), but support for new catalogs is added as they become available. Catalog files must be stored off-line in the computer. Astrometric match is performed similarly to the mutual image alignment except that one set of triangles is artificially created from tangentially projected star coordinates from the catalog files. Users can define the number of brightest stars used to create triangles for each image and catalog plate. If a matching triangle is found, a transformation matrix is calculated, and tested on all stars used to create triangles. If $50\%$ or more stars are matched, another transformation matrix is calculated from all matched stars. For astrometry purposes, every detected star in the image is searched in the catalog and if the corresponding catalog star is found, the pair is recorded. Then a final transformation matrix is calculated from all matched stars.

\subsubsection{Field curvature}

Wide field optical systems often suffer from less than ideal tangential projection of stars to the imaging area of the detector. Depending on the field of view, differences between ideal tangential projection and the actual centroid of the projected star can be tens of arcseconds. In such case, matching the catalog to stars in the image is not possible. To overcome this problem, SIPS allows for the definition of field curvature. The correction is based on two two-dimensional $3^\mathrm{rd}$-order polynomials describing the difference between actual and ideal tangential projections in the $x$ and $y$ directions. SIPS can also calculate these polynomials from matched pairs of image and catalog stars using the least-square method (2D $3^\mathrm{rd}$-order polynomial regression). To solve the challenge of providing an approximate initial solution, SIPS contains a tool allowing the user to manually mark image-catalog star pairs, which is then used to calculate field curvature description. With the correction solved, the polynomial coefficients can be saved into a file and later automatically re-used for astrometric solution.

\subsubsection{Photometry}

Aperture photometry is calculated for $10$ apertures. Default aperture radii $r_a$ range from 2 to approx. 14 pixels, which should fit virtually all combinations of detectors and optics used by amateur astronomers. Still, users are free to define own apertures.  In addition to $10$ predefined apertures, SIPS calculates photometry also for aperture radius automatically determined during the process of refining of bitmap of stellar images as is described above. To be usable for photometry measurement, aperture radii of every star must be the same for all images in the set. There is also the lower limit for the automatic aperture in the case the calculation results in too small value. The smallest auto aperture default value is set to two pixels, but this parameter is also adjustable by the user. To find the automatic aperture for a set of images, SIPS sorts the images according to the radius of the automatically determined aperture for every star. Then it chooses the third largest aperture. This empirical rule is intended to select the largest aperture to cover the whole star when, for example, seeing changes during an observation session. At the same time, the algorithm removes up to two extreme values caused, for example, by tracking glitches or similar random events.

Mean value $F_\text{b}$ and standard deviation $\sigma_\text{b}$ of the background flux are calculated from pixels in a ring characterized by two user-specified radii. The inner ring radius is always larger than the biggest aperture used to calculate photometry. SIPS supports two methods of background mean and standard deviation calculation. The first method relies on the previously calculated bitmask, which allows skipping pixels belonging to stars. Background statistics are then calculated only from pixels between the two radii, which are also not flagged in the corresponding bitmask. To eliminate random and mostly single-pixel artifacts not detected as stars, all pixels above or below the mean plus or minus 3-times the standard deviation are rejected as outliers. The newly calculated mean and standard deviation are then taken as valid values for background. While this method is fast and reliable for sparse star fields, it is not robust enough for dense fields with many stars within the background ring or when artifacts (e.g., diffraction spike from a nearby bright star, sub-atomic particle, or satellite trace) significantly affect the background ring. As a second option, SIPS implements also a slower but more robust mean calculation using the Hampel influence function \citep{Hampel2005-hv}.

Flux for photometric aperture $F_a$ with radius $r_a$ is calculated as a sum of pixel values above the mean level of the background ring as
\begin{equation}
    F_a = \sum_i f_i \text{ADU}_i - N_a F_\text{b},
\end{equation}
where the sum proceeds over all pixels at least partially contained within the aperture, ADU$_i$ is the flux (Analog-to-Digital Unit) of the pixel at coordinates $(p_x,p_y)$, $f_i$ is the fraction of the pixel contained in the aperture, and $N_a = \sum_i f_i$ is the total number of pixels in the aperture. The fraction of a pixel contained within the aperture is calculated from the Euclidean distance $r$ of the star pixel from the star centroid at $(c_x,c_y)$, $r^2 = (p_x-c_x)^2 + (p_y-c_y)^2$, as 
\begin{equation}
    {f}_i = \begin{cases}
    1, & r \le r_a -0.5\\
    r_a + 0.5 - r, & r_a-0.5 < r \le r_a + 0.5\\
    0, & \text{otherwise}.
    \end{cases}
\end{equation}
Flux uncertainty is calculated as
\begin{equation}
\sigma_{a} = \sqrt{\sum_i f_i \text{ADU}_i+N_a {\sigma_\text{b}}^2}.
\end{equation}

SIPS calculates only relative photometry, where fluxes from the variable star $F_{a,V}$ are compared to fluxes from at least one comparison star $F_{a,C}$ as
\begin{equation}
\mathcal{F} = \frac{F_{a,V}}{F_{a,C}}.
\end{equation}
The flux ratio uncertainty is
\begin{equation}
\sigma_F = \mathcal{F} \left(\frac{\sigma_{a,V}}{F_{a,V}} + \frac{\sigma_{a,C}}{F_{a,C}} \right).
\end{equation}
User can specify one or more comparison stars. For more than one comparison star the result is averaged over the ensemble.

Alternatively, the result of photometry are instrumental magnitudes defined as
\begin{equation}
m = -2.5 \log_{10} \mathcal{F},
\end{equation}
with uncertainty calculated with
\begin{equation}
\sigma_m = \frac{2.5}{\ln 10} \sqrt{ \left( \frac{\sigma_{a,V}}{F_{a,V}} \right)^2 + \left( \frac{\sigma_{a,C}}{F_{a,C}} \right)^2 }.
\end{equation}

Finally, SIPS requires all pixels within the photometric aperture to lie below the saturation value. If any pixel exceeds saturation, then the whole photometric point is invalidated, because it is not possible to determine the amount of lost flux. The saturation value is by default set to $(2^{16}-1) = 65535$, which is valid for cameras with 16-bit Analog-to-Digital Converter (ADC). However, many modern CMOS cameras are equipped with only 12-bit or 14-bit ADC. In such situations, SIPS relies on a FITS header keyword DATAMAX, which provides the saturation value.

\subsubsection{User interface for photometry processing}

Full description of the SIPS photometry package user interface is beyond the scope of this Appendix. In general, SIPS strives  to enhance the efficiency of photometry processing and reduce user effort to process many wide-field images containing tens or hundreds of objects of interest and to save the resulting light curves quickly for further processing. 

Here, we mention in more detail two features. First, the photometry task tool plans the complete processing of the observing series. When a task for a particular series is defined, photometry processing can be performed by a single mouse click. Task describes which files from which folder should be processed, which calibration files should be used, and which field description is to be loaded after processing is complete. The photometry task can rely on SIPS default parameters for star search, alignment, astrometry, and photometry reduction, but it is also possible to override default values with task-specific ones, including field curvature description file and many other options.

Second, SIPS provides the field description tool to easily mark all the stars of interest (variable and comparison stars) within the field of view. The field description file consists of two parts. The first part is a list of all named stars within a field of view. Names are user-defined identifiers of variable stars, comparison, and possibly check stars. If the named star is paired with a catalog star on the basis of its equatorial coordinates, catalog identification and coordinates are automatically included as well. The second part consists of N-tuples of a variable star name, followed by one or more comparison star names and a check star name. At last, a variable and one comparison stars are mandatory; more comparison stars and check stars are optional. 

SIPS graphical user interface allows immediate plotting of light curves of variable stars defined in the field description as well as storing photometry protocol (text file), optionally accompanied by an image, created from the selected frame with marked variable, comparison, and check stars and plotted used apertures. This facilitates quick visualization of the large number of light curves.

\subsubsection{Identification of variable objects}

SIPS can identify stars, which significantly change their instrumental magnitude during the observing run. This feature can be used to search for new variable stars or exoplanetary transits. Variable star candidates can be identified using three methods:
\begin{enumerate}
    \item {The simplest method calculates standard deviation of instrumental magnitudes of each star over all images in a series. SIPS then plots mean instrumental magnitude against the standard deviation, which gives the typical ``hockey stick'' shaped group of points. Variable stars with higher standard deviation than non-variable stars of similar brightness are located above the main locus of points. User can easily manually investigate light curve of any star.}
    \item{SIPS also implements a method inspired by Eq.~(3) of \citet{Jaimes2013}, which is based on previous work of \citet{tamuz06} and \citet{arellano12}. This method helps to reduce inefficiency of method 1 in fields with high number of stars and correspondingly also with a high number of falsely positive candidates.}
    \item{Finally, SIPS uses a simple three-layer neural network with sigmoid transfer function. Each light curve is equidistantly resampled to 512 points using linear interpolation and the amplitude is normalized to range from 0 to 1. The result is then supplied to the network input layer consisting of 512 neurons. The middle layer contains 256 neurons and the output layer consists of single neuron, which gives the final probability that the light curve corresponds to a variable star. The network is trained on a set of approximately 5000 examples of both variable and non-variable stars and the network configuration is included in the SIPS package. Variable stars in the training set cover the majority of common types, including detached, semi-detached and contact eclipsing binaries and pulsating stars of various types. Users are free to create their own training sets and re-train the network to better recognize the typical light curve shapes resulting from sampling cadence, angular resolution, image quality etc. Performance of the neural network detection of variable stars depends on the numerous factors such as relative percentage of artifacts and overall quality of data. However, for data sets similar to the ones used to train the network, the results are typically superior to the methods 1 or 2. Owing to the fact that the input light curve amplitudes are normalized, the neural network response depends virtually on the light curve shape only. Compared to first two methods, which rely on statistical properties of the series of brightness measurements, the neural network response is not affected by absolute transit depth or the ratio between transit depth and brightness deviation. This makes the neural network especially sensitive to variable stars with shallow transits.}
\end{enumerate}

\subsection{SILICUPS (SImple LIght CUrve Processing System)}

SILICUPS software package is designed to gain an overview of observed data and maintain light curves of many objects acquired through many nights in one place. Here, we focus only on the description of SILICUPS functions relevant to this work, specifically,  phenomenological fitting of minima timings and entire light curves, which were used to disentangle signals from doubly eclipsing binaries and build $O-C$ diagram shown in Figure~\ref{fig:oc_orig}.

SILICUPS works with fields, consisting of multiple objects. Every field is stored as a separate description file and SILICUPS handles one field at a time. It is upon the user if the file contains only objects from one field of view of the used camera/telescope or objects spread over a large portion of the sky. However, some advanced functions, like checking the presence of the field objects in the AAVSO VSX database, work properly only on fields with angularly nearby objects.

Every object within a field has an associated individual series of photometric points. Series are read from photometric report text files, which include at minimum JD, magnitude or flux, and error (uncertainty). Such files are generated for instance by SIPS or other photometry software or they can be created from publicly available photometric databases (for instance from TESS data). Any line of the photometric protocol file, which does not start with a number, is considered a comment line. Comment lines may contain keywords followed by values, which define metadata (star identification, coordinates, catalog cross-id, used filter or time standard -- heliocentric or geocentric, etc.).

To avoid manual entry of each report file, which could be very time-consuming especially when single observing session easily produces more than 100 new reports, SILICUPS relies on a definition of rules to find all relevant protocol files in the computer file system. Such rules consist of one or more protocol file names, possibly containing wild-card characters, folder name keywords, and others. SILICUPS is then able to automatically update its light-curve database without user intervention.

SILICUPS contains two methods of determining variable star periods. The first method relies on known  minima timings and uses a brute-force algorithm to find periods for which minima instances align with defined precision. This method is suitable especially for eclipsing binaries, where at last three minima instances are known. The second method is based on frequency analysis \citep{Czerny1996} and is suitable especially for physical pulsating stars.

Minima timings within individual series can be determined by fitting with the phenomenological model defined by \citet{Mikulasek2015}, which can be optionally extended with a trend (slope) to represent instrumental effects. SILICUPS can fit entire phased light curves using a combination of functions from \citet{Mikulasek2015} and a second-order trigonometric polynomial to represent out of eclipse variations. The fits are performed using  least squares implemented in {\tt cmpfit} \citep{more78,markwardt09}. Uncertainties are estimated using a bootstrapping algorithm. Any parameter can be fixed to a predefined value using a graphical interface.

When the phenomenological model is defined for the object, SILICUPS allows exporting of the time series with the model subtracted from the original data. This feature allows to disentangle light curves containing signals from multiple eclipsing binaries and create light curves of individual components. $O-C$ analysis of separated components then allows studying of LTTE and other effects.

Finally, we point out that functions of SIPS and SILICUPS are continuously evolving also in response to user requests. For example, we are now experimenting with online data synchronization of data between different users to enable efficient collaboration.

\section{Ground-based light curves}
\label{app:ground}

\begin{table}
\caption{Ground-based photometry. Full version is available in the on-line version.}
\label{tab:ground}
    \centering
    \begin{tabular}{ccc}
    \hline\hline
    HJD - 2 450 000 & Magnitude & Uncertainty\\
    \hline
    5957.352240 & 0.9643 & 0.0069 \\
5957.355290 & 0.9727 & 0.0069 \\
5957.358350 & 0.9634 & 0.0071 \\
5957.361400 & 0.9548 & 0.0072 \\
5957.364460 & 0.9617 & 0.0072 \\
5957.367520 & 0.9632 & 0.0073 \\
5957.370570 & 0.9756 & 0.0073 \\
5957.373630 & 0.9667 & 0.0073 \\
5957.376680 & 0.9720 & 0.0072 \\
5957.379740 & 0.9678 & 0.0071 \\
5957.382800 & 0.9779 & 0.0072 \\
5957.385850 & 0.9803 & 0.0071 \\
5957.388910 & 0.9852 & 0.0072 \\
5957.391970 & 0.9835 & 0.0070 \\
5957.395020 & 0.9837 & 0.0069 \\

    \hline
    \end{tabular}
\end{table}

\begin{figure*}
    \centering
    \includegraphics[width=0.99\textwidth]{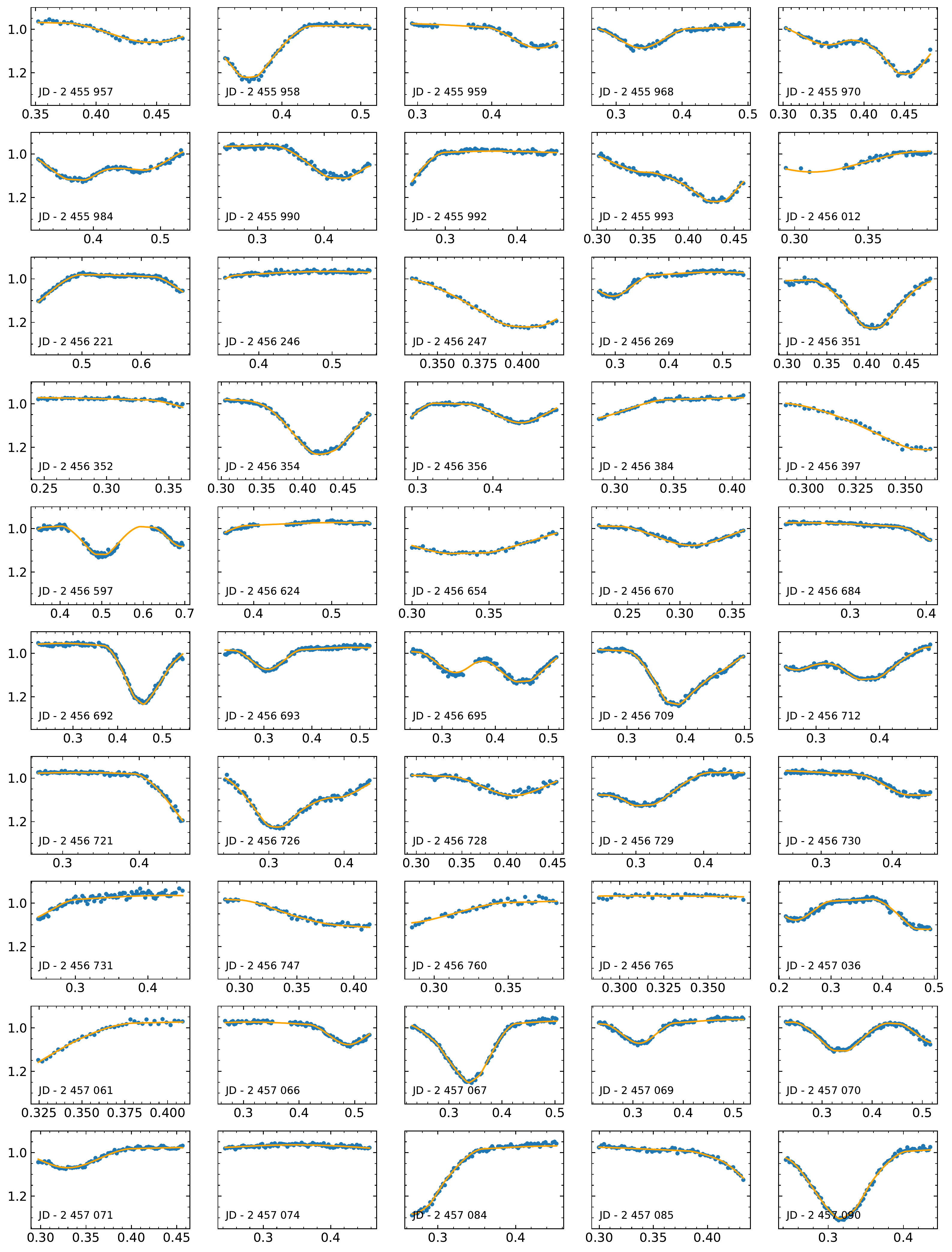}
    \caption{Fits of ground-based light curves, part 1}
    \label{fig:ground_lc1}
\end{figure*}

\begin{figure*}
    \centering
    \includegraphics[width=0.99\textwidth]{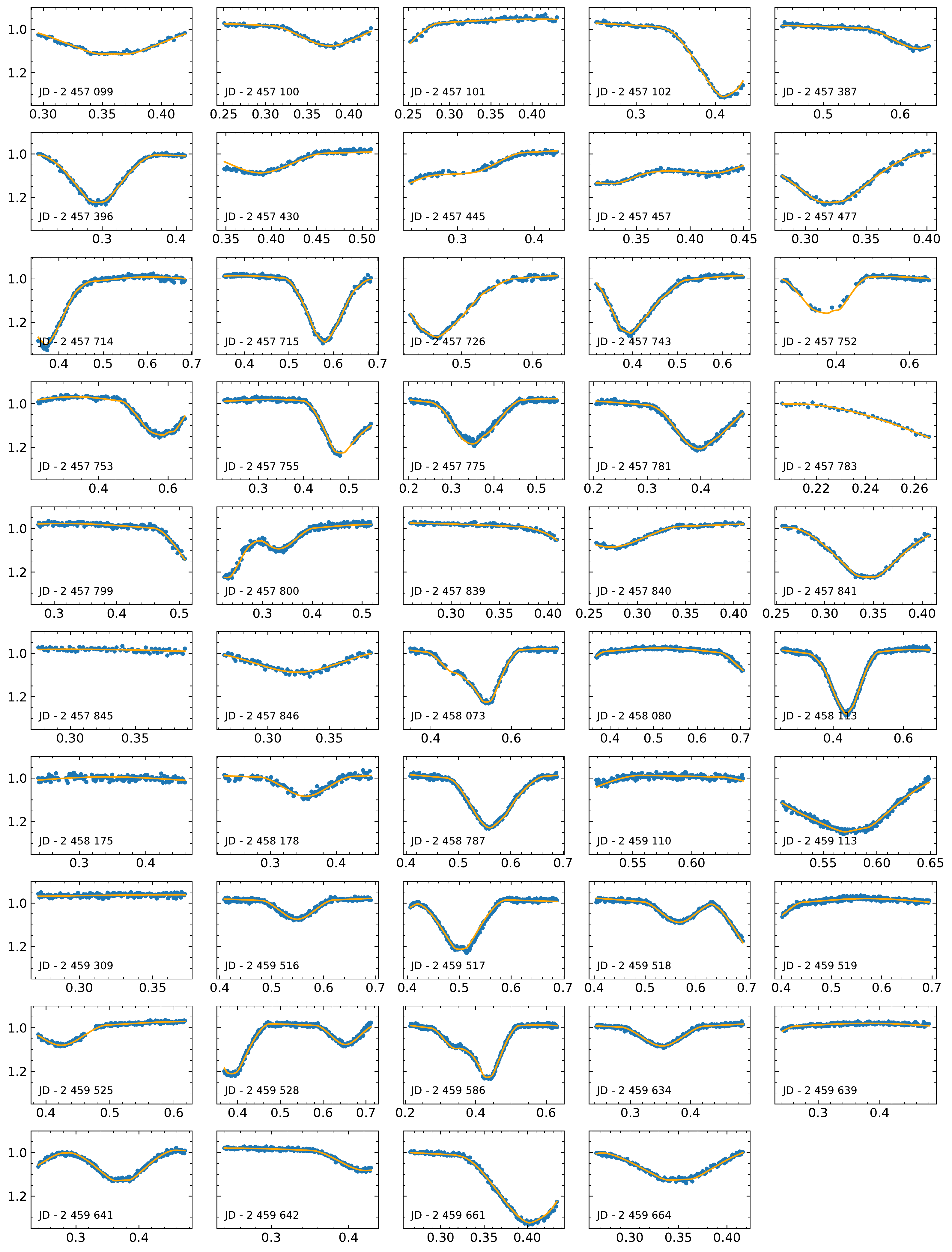}
    \caption{Fits of ground-based light curves, part 2}
    \label{fig:ground_lc2}
\end{figure*}

In Table~\ref{tab:ground}, we present our ground-based photometry of CzeV343. Our dataset includes the measurements already published by \citet{cagas12}, but the data differ in detail due to changes in image processing and we thus reproduce them here. Light curves from individual observing nights along with the best-fit model are shown in Figures~\ref{fig:ground_lc1} and \ref{fig:ground_lc2}.

\section{TESS light curves}
\label{app:tess}

\begin{figure*}
    \centering
    \includegraphics[width=\textwidth]{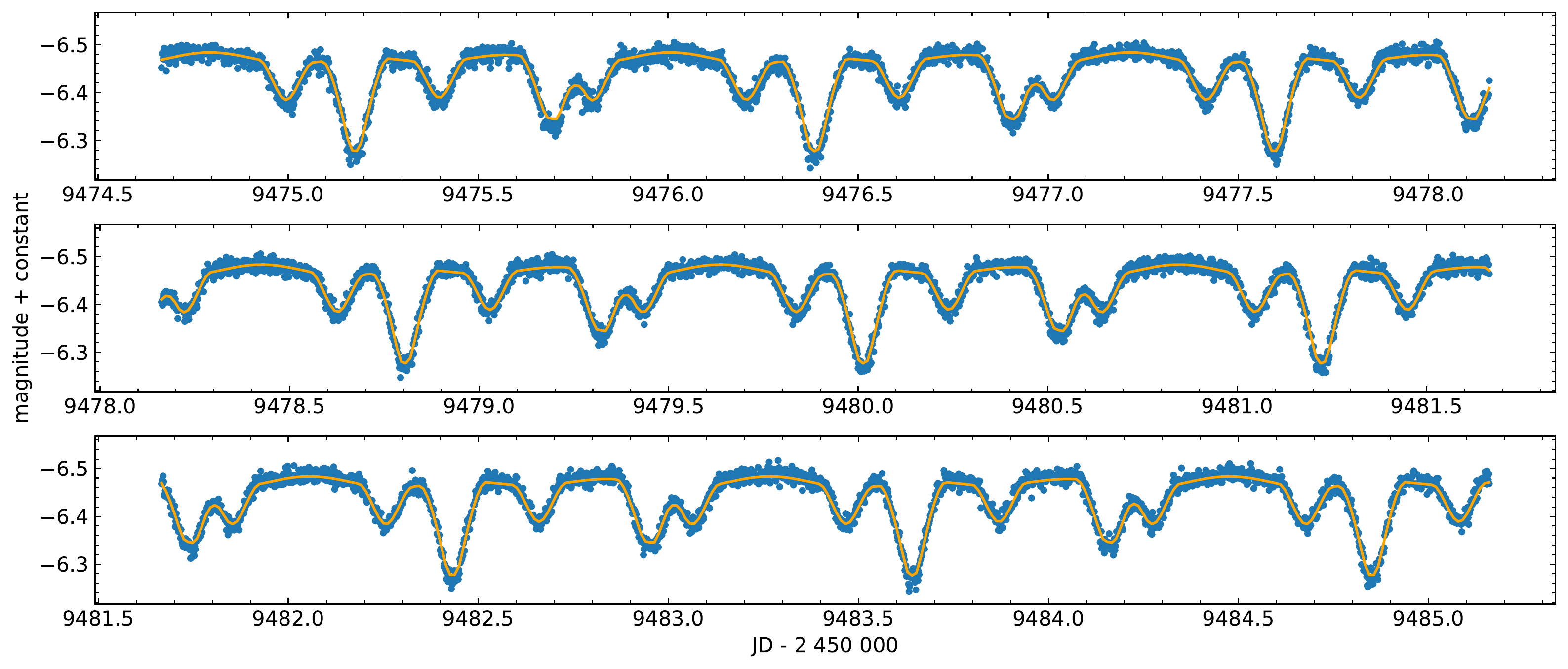}\\
    \includegraphics[width=\textwidth]{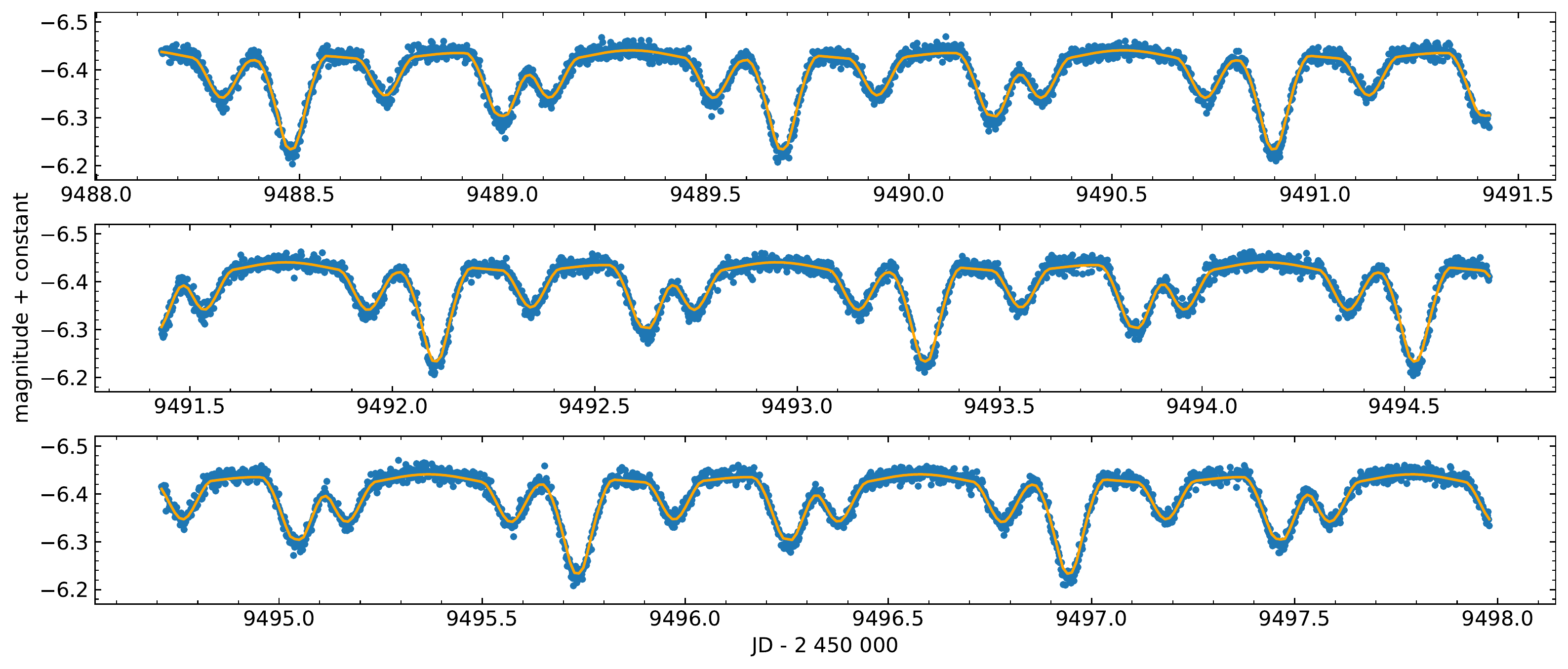}\\
    \includegraphics[width=\textwidth]{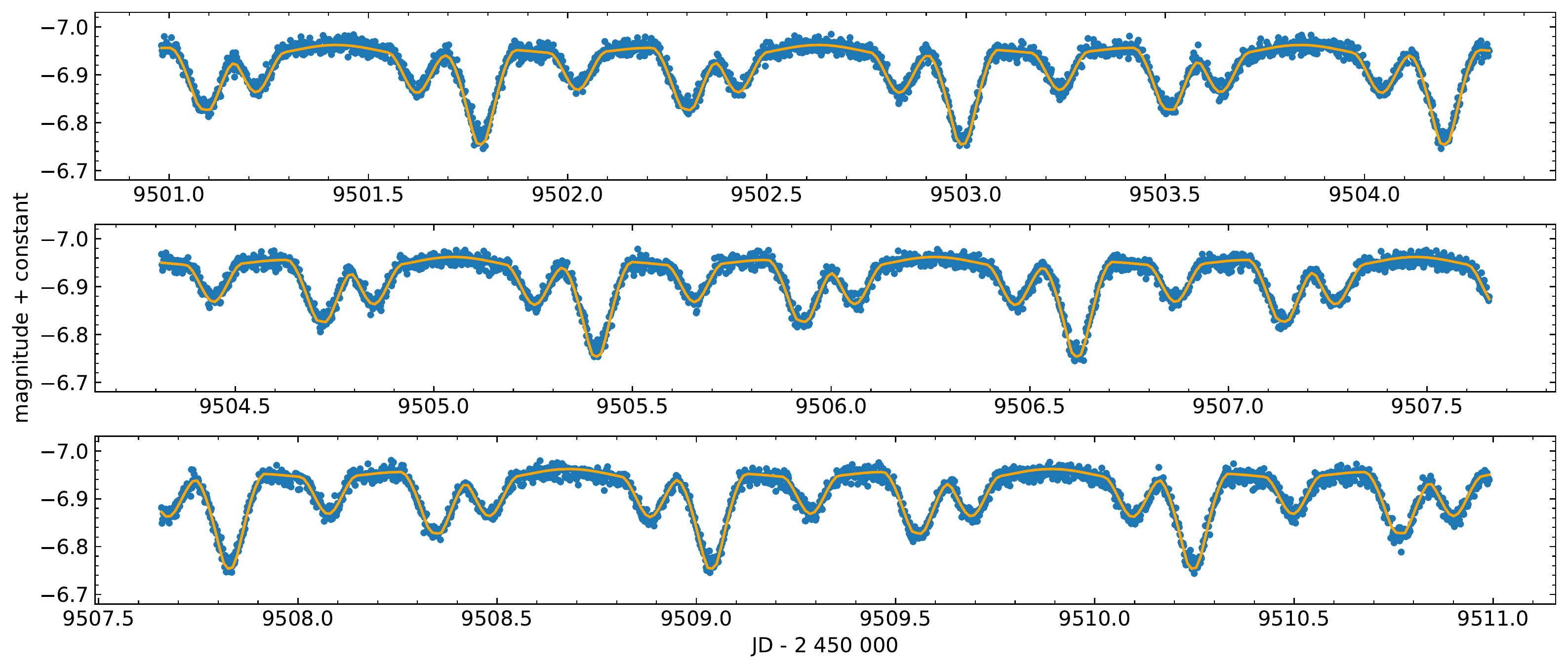}
    \caption{Fits of TESS light curves, part 1}
    \label{fig:tess_lc1}
\end{figure*}

\begin{figure*}
    \centering
    \includegraphics[width=\textwidth]{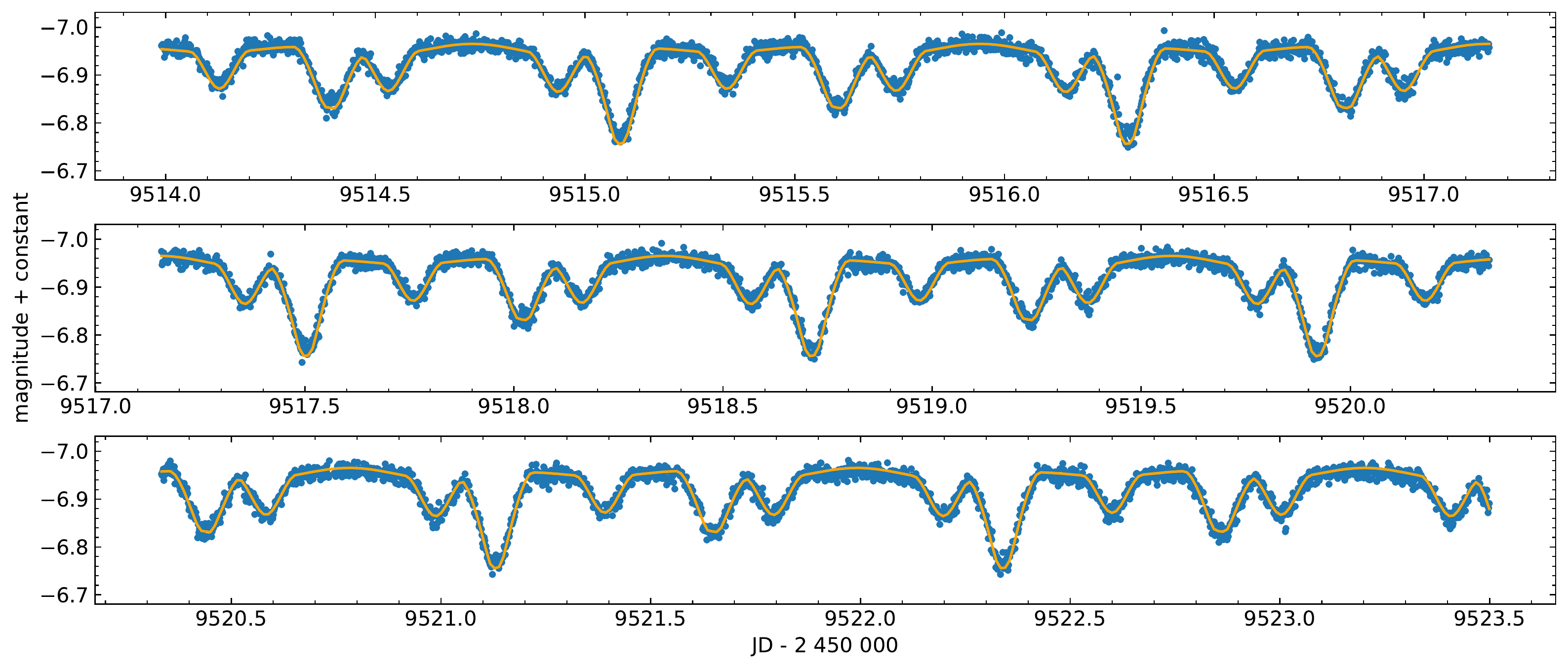}\\
    \includegraphics[width=\textwidth]{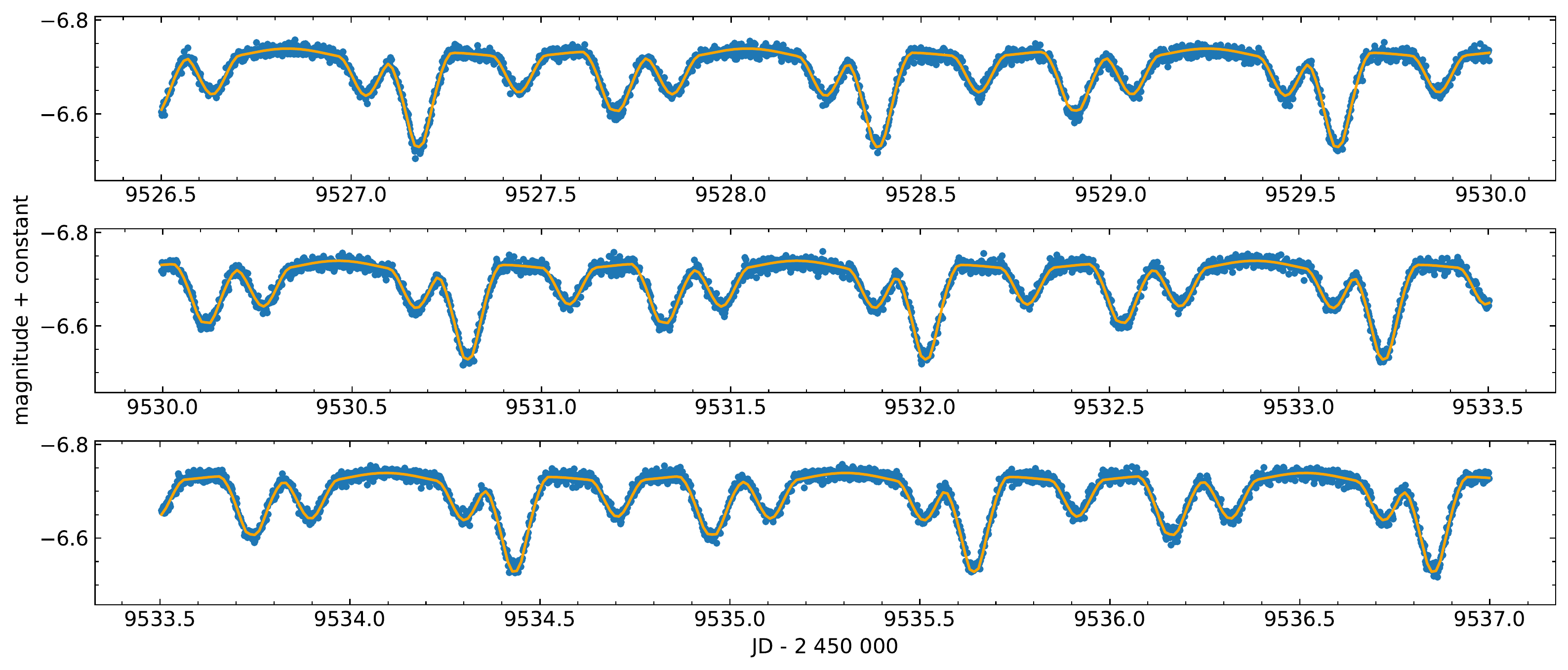}\\
    \includegraphics[width=\textwidth]{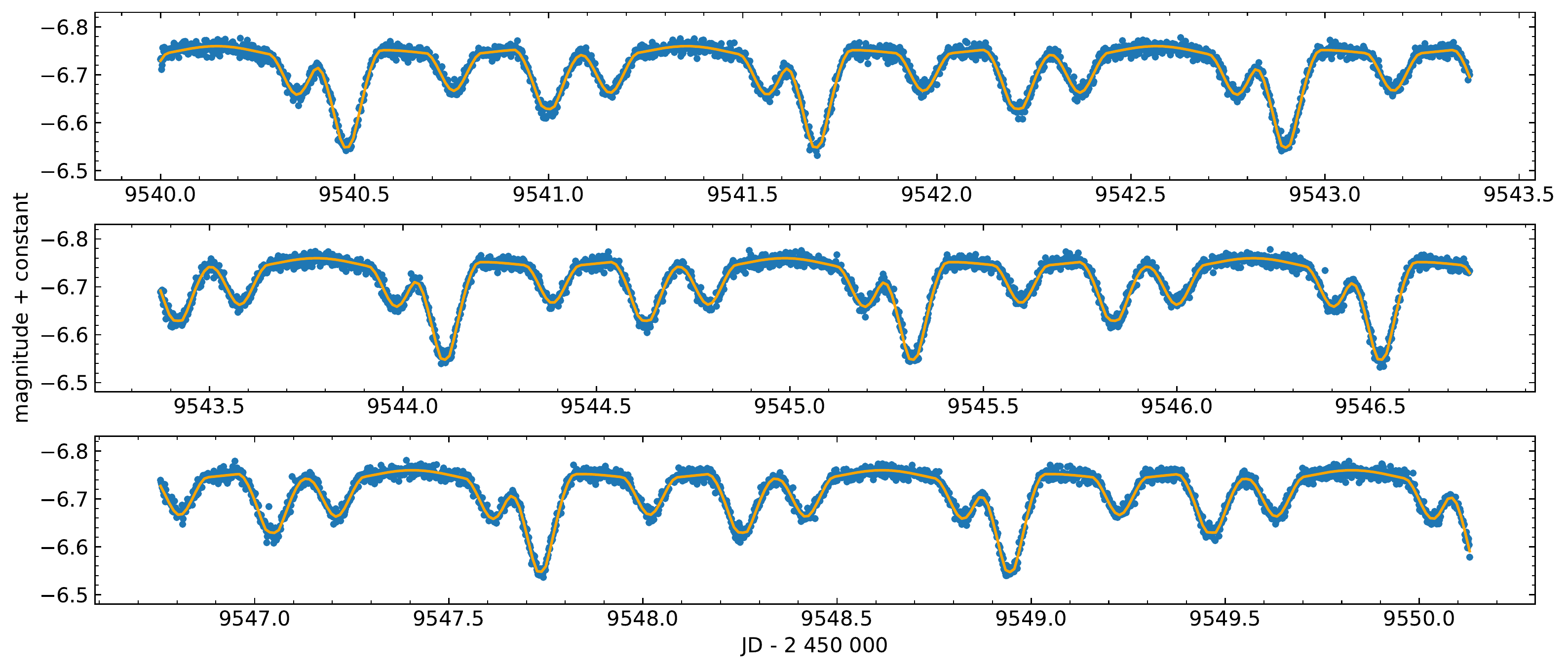}
    \caption{Fits of TESS light curves, part 2}
    \label{fig:tess_lc2}
\end{figure*}

\begin{figure*}
    \centering
    \includegraphics[width=\textwidth]{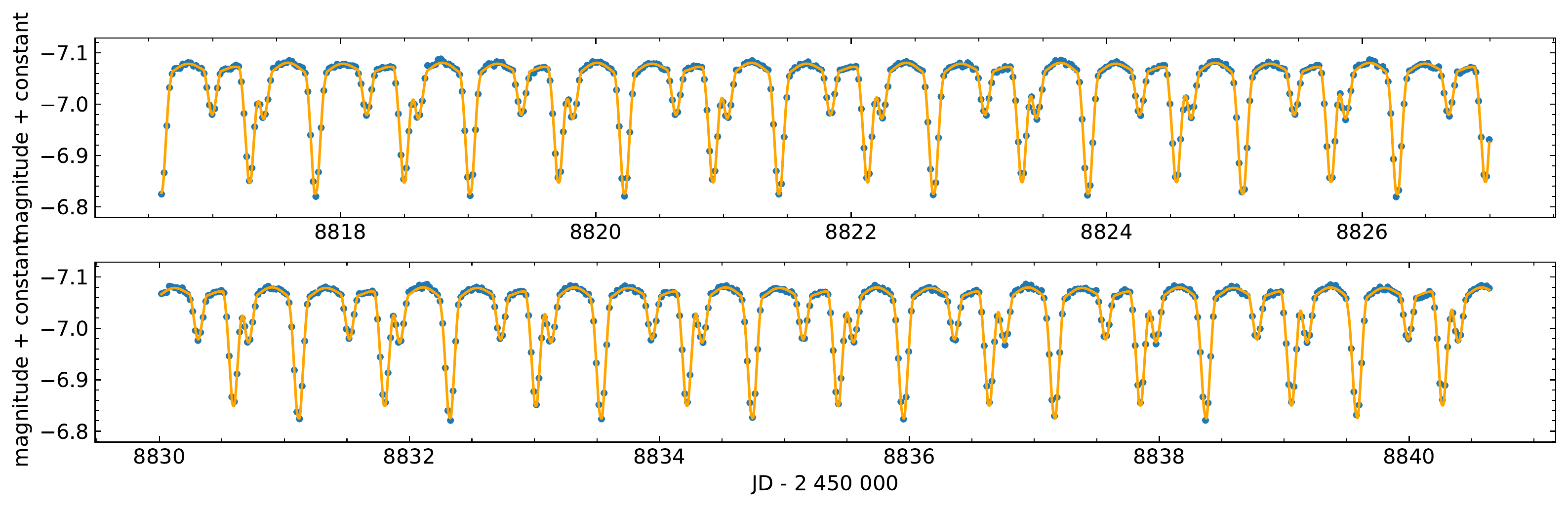}\\
    \caption{Fits of TESS FFI light curves.}
    \label{fig:ffi}
\end{figure*}

We show TESS 2-minute light curves along with their fits in Figures~\ref{fig:tess_lc1} and \ref{fig:tess_lc2}. In Figure~\ref{fig:ffi}, we show the same for TESS FFI data with best-fit PCA vectors already subtracted.

\section{Posterior distributions of parameters}
\label{app:posteriors}

\begin{figure*}
    \centering
    \includegraphics[width=\textwidth]{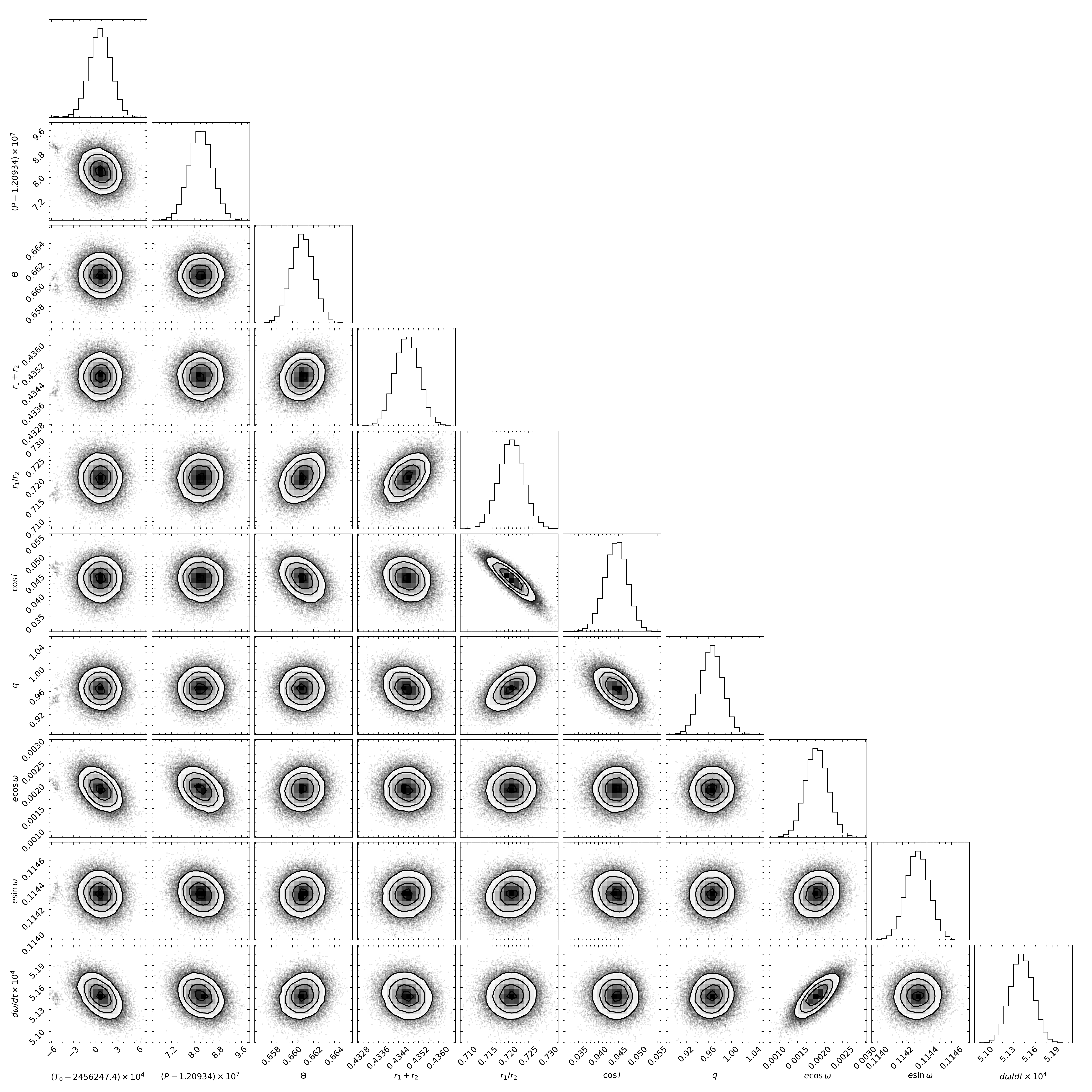}
    \caption{Posterior distributions of parameters of binary $A$.}
    \label{fig:corner_a}
\end{figure*}

\begin{figure*}
    \centering
    \includegraphics[width=\textwidth]{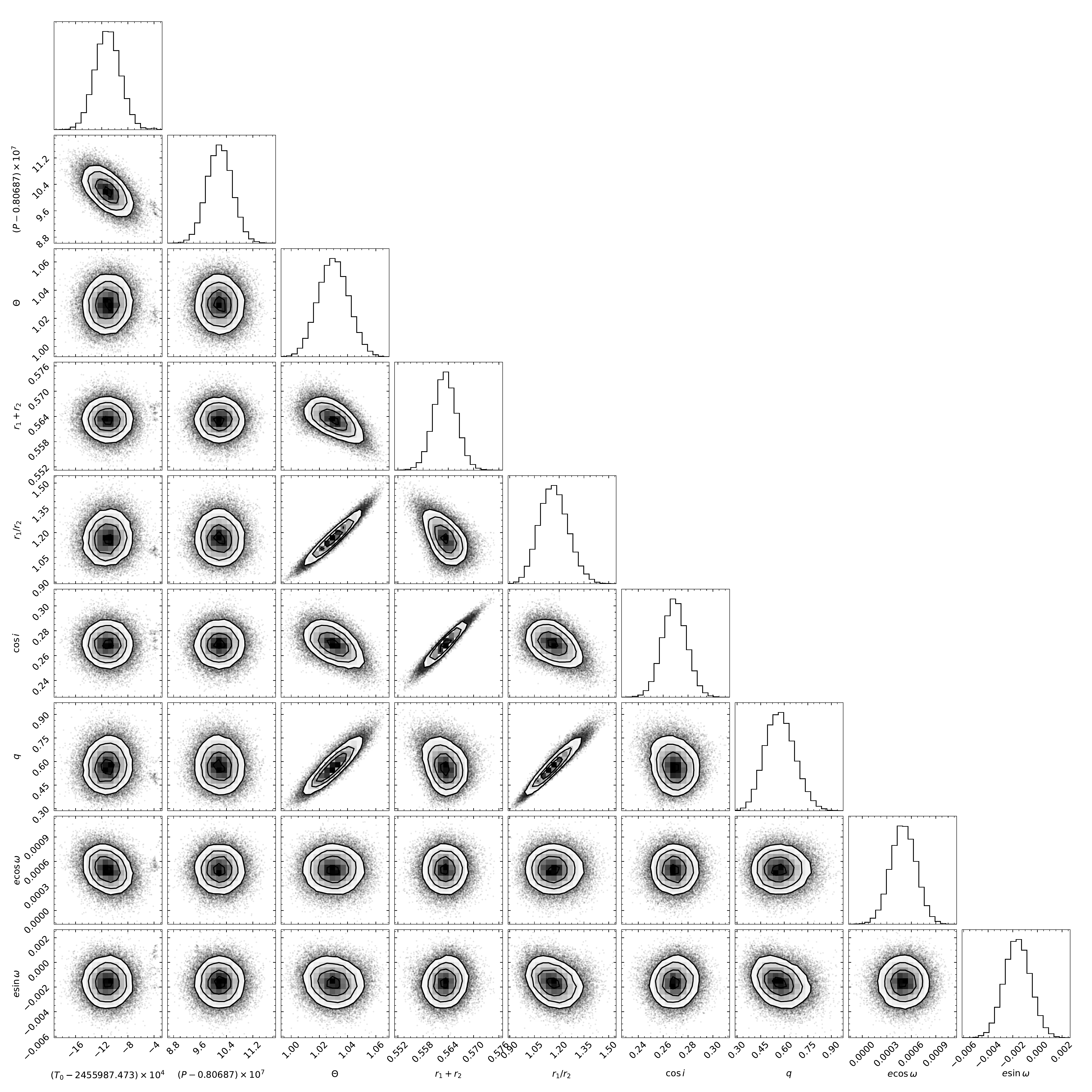}
    \caption{Posterior distributions of parameters of binary $B$.}
    \label{fig:corner_b}
\end{figure*}

\begin{figure*}
    \centering
    \includegraphics[width=0.8\textwidth]{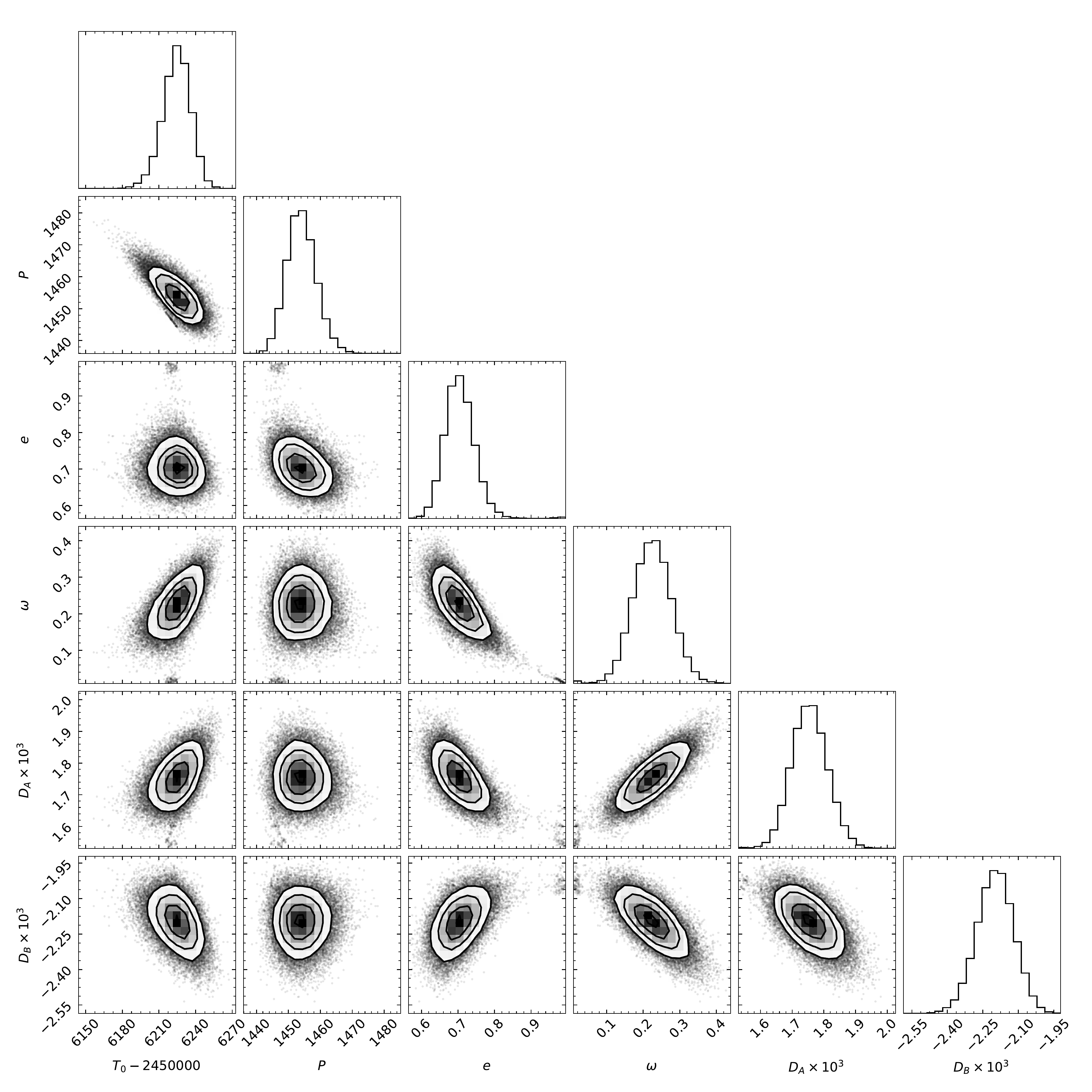}
    \caption{Posterior distributions of parameters of the mutual orbit.}
    \label{fig:corner_c}
\end{figure*}

\begin{figure*}
    \centering
    \includegraphics[width=0.8\textwidth]{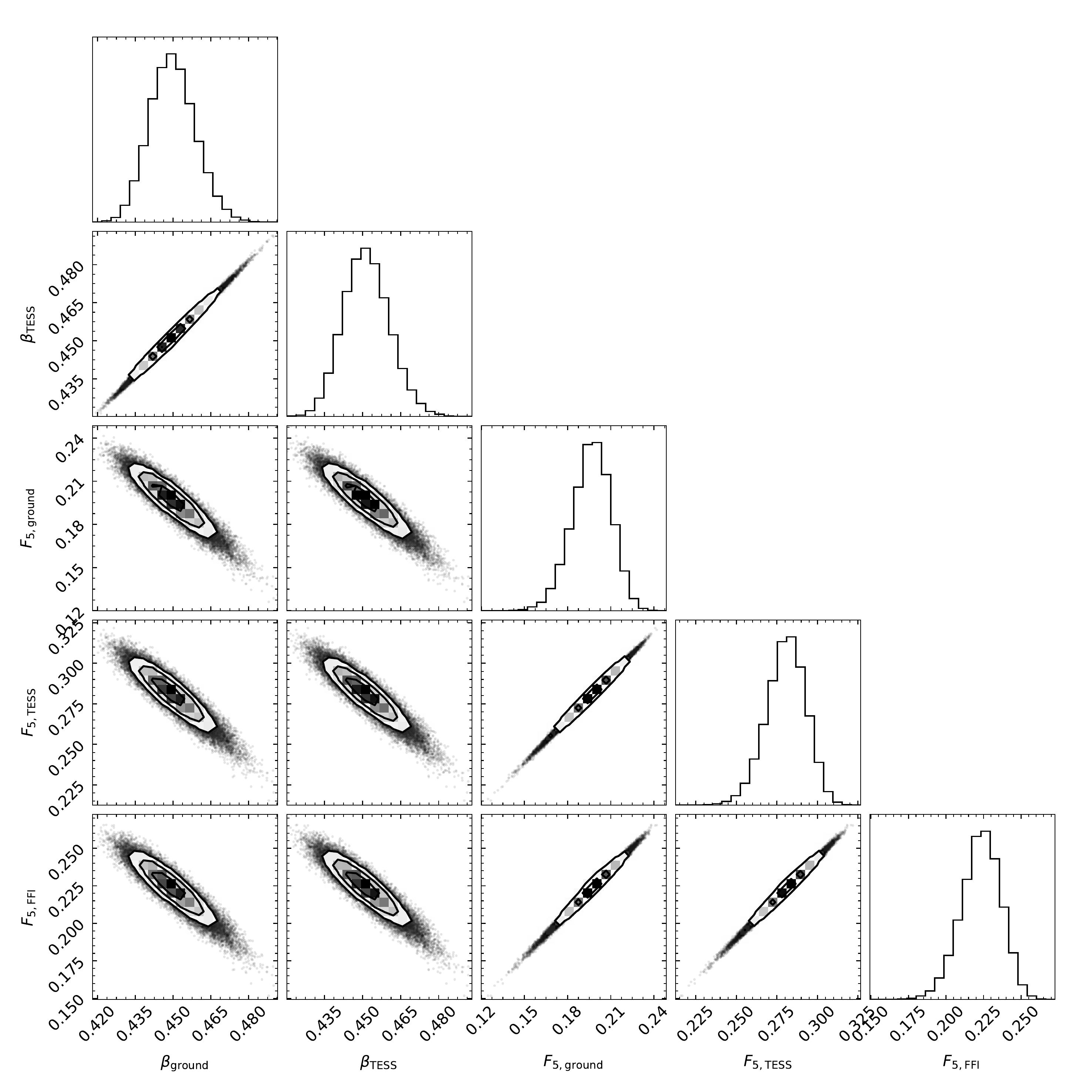}
    \caption{Posterior distribution of the auxilliary parameters of the photometric model.}
    \label{fig:corner_aux}
\end{figure*}

Figures~\ref{fig:corner_a}, \ref{fig:corner_b}, \ref{fig:corner_c}, and \ref{fig:corner_aux} show posterior distributions of the parameters of our model. The figures were made with package {\tt corner} \citep{corner}.

\end{appendix}

\end{document}